\RequirePackage{fix-cm}
\documentclass[twocolumn]{svjour3}          
\smartqed  
\usepackage{graphicx}
\usepackage{mathptmx}      
\usepackage{latexsym}
\usepackage{amsmath,amssymb} 
\usepackage[dvipsnames]{xcolor}
\usepackage[utf8]{inputenc}
\begin{document}

\title{Exploration of the high-redshift universe enabled by THESEUS 
}


\author{
N. R. Tanvir     \and
E. Le Floc'h  \and
L. Christensen \and
J. Caruana \and
R. Salvaterra \and
G. Ghirlanda \and
B. Ciardi \and
U. Maio \and
V. D'Odorico \and
E. Piedipalumbo \and
S. Campana \and
P. Noterdaeme \and
L. Graziani \and
L. Amati \and
Z. Bagoly \and
L. G. Bal\'azs \and
S. Basa \and
E. Behar \and
E. Bozzo \and 
A. De Cia \and
M. Della Valle \and
M. De Pasquale \and
F. Frontera \and
A. Gomboc \and
D. G\"otz \and
I. Horvath \and
R. Hudec \and
S. Mereghetti \and
P. T. O'Brien \and
J. P. Osborne \and
S. Paltani \and
P. Rosati \and
O. Sergijenko \and
E. R. Stanway \and
D. Sz\'ecsi \and
L. V. To\'th \and
Y. Urata \and
S. Vergani \and
S. Zane 
}


\institute{N. R. Tanvir \at
              School of Physics and Astronomy, University of Leicester, University Road, Leicester. LE1 7RH. United Kingdom \\
              Tel.: +44-116-2231217\\
              \email{nrt3@le.ac.uk}           
           \and
           E. Le Floc'h \at
             AIM, CEA-Irfu/DAp, CNRS, Université Paris-Saclay,
             F-91191 Gif-sur-Yvette, France
             \and
             L. Christensen \at Niels Bohr Institute,
             University of Copenhagen, Jagtvej 128, 2200 Copenhagen N, Denmark
             \and
             J. Caruana \at Department of Physics and Institute of Space Sciences and Astronomy, University of Malta, Msida MSD~2080, Malta
             \and
             R. Salvaterra \at                INAF-Istituto di Astrofisica Spaziale e Fisica cosmica,               via Alfonso Corti 12, 20133 Milano, Italy
             \and
             G. Ghirlanda \at INAF-Osservatorio astronomico di Brera, Via Bianchi 46, I-23807, Merate (LC), Italy
             \and
             B. Ciardi \at Max Planck Institute for Astrophysics, Karl-Schwarzschild-Str. 1, 85741 Garching, Germany
             \and
             U. Maio, V. D'Odorico \at INAF, Observatory of Trieste, via G. Tiepolo 11, 34143 Trieste, Italy 
             \and
                         E. Piedipalumbo \at Dipartimento di Fisica, Universit\`a degli Studi di Napoli Federico II, 80126 Naples, Italy ; I.N.F.N., Sez. di Napoli, Compl. Univ. Monte S. Angelo, via Cinthia, 80126 - Napoli, Italy
             \and
             S. Campana \at INAF-Osservatorio astronomico di Brera, Via Bianchi 46, I-23807, Merate (LC), Italy
             \and
             P. Noterdaeme \at Institut d'Astrophysique de Paris, CNRS-SU, UMR\,7095, 98bis bd Arago, 75014 Paris, France 
             \and
             L. Graziani \at Dipartimento di Fisica, ``Sapienza'' Universit\`a di Roma, Piazzale Aldo Moro 5, 00185 Roma, Italy
             \and
             L. Amati \at INAF/OAS-Bologna, via P. Gobetti 101, I-40129 Bologna, Italy
             \and
             Z. Bagoly \at E\"ov\"os University, H-1117 Budapest, Hungary
             \and
              L.~G. Bal\'azs \at CSFK Konkoly Observatory, Budapest,
                         E\"ov\"os University, Dept. of Astronomy, Budapest, Hungary
             \and
             S. Basa \at Aix Marseille Univ, CNRS, CNES, LAM, Marseille, France
           \and
            E. Behar \at Department of Physics, Technion, Israel
             \and
            E. Bozzo, S. Paltani \at Department of Astronomy, University of Geneva, ch. d'\'Ecogia 16, 1290 Versoix, Switzerland      
             \and 
            A. De Cia \at Department of Astronomy, University of Geneva, Chemin Pegasi 51, 1290 Versoix, Switzerland
             \and
             M. Della Valle \at INAF-Capodimonte Observatory, Salita Moiariello 16, 80131, Napoli, Italy
             \and
            M. De Pasquale \at Department of Astronomy and Space Sciences, Istanbul University, Beyazıt 34119, Istanbul, Turkey
             \and
             F. Frontera \at Dipartimento di Fisica e Scienze della Terra, Universit\`a di Ferrara, Via Saragat 1, I-44122 Ferrara, Italy
             \and
             A. Gomboc \at Center for Astrophysics and Cosmology, University of Nova Gorica, Vipavska 13, 5000 Nova Gorica, Slovenia
             \and
             D. G\"otz \at IRFU/D\'epartement d'Astrophysique, CEA, Universit\'e Paris-Saclay, F-91191 Gif-sur-Yvette, France
             \and
            I. Horvath \at University of Public Service, Budapest, Hungary
             \and
            R. Hudec \at Czech Technical University in Prague, Faculty of Electrical Engineering, Prague, Czech Republic;
 Astronomical Institute, Czech Academy of Sciences, Ondrejov, Czech Republic; Kazan Federal University, Kazan, Russian Federation
 \and
 S. Mereghetti \at INAF / IASF-Milano, via A. Corti 12, I-20133 Milano, Italy
 \and
P. T. O'Brien, J. P. Osborne \at
              School of Physics and Astronomy, University of Leicester, University Road, Leicester. LE1 7RH. United Kingdom
\and
P. Rosati,
\at Department of Physics and Earth Sciences, University of Ferrara, Ferrara, Italy
\and
            O. Sergijenko \at Astronomical Observatory of Taras Shevchenko, National University of Kyiv, Observatorna str., 3, Kyiv, 04053, Ukraine; Main Astronomical Observatory of the National Academy of Sciences of Ukraine, Zabolotnoho str., 27, Kyiv, 03680, Ukraine 
            \and
E. R. Stanway \at University of Warwick, Physical Sciences, Gibbet Hill Road, Coventry, CV4 7AL, United Kingdom
            \and
             D. Sz\'ecsi \at Institute of Astronomy, Nicolaus Copernicus University, 
             87-100 Toru\'{n}, 
             Poland ; I. Physikalisches Institut, Universit\"at zu K\"oln, D-50937 Cologne, Germany
             \and
L. V. To\'th \at Department of Astronomy of the  E\"otv\"os Lor\'and University, P\'azm\'any P\'eter s\'et\'any 1/A, H-1117 Budapest, Hungary
\and
             Y. Urata \at Institute of Astronomy, National Central University, Chung-Li 32054, Taiwan
\and 
S. D. Vergani \at GEPI, Observatoire de Paris, PSL University, CNRS, Place Jules Janssen, 92190 Meudon, France
\and
            S. Zane \at Mullard Space Science Laboratory, University College London, Holmbury St Mary, Dorking,  RH56NT,  UK
}

\date{Received: date / Accepted: date}

\maketitle

\begin{abstract}
At peak, long-duration gamma-ray bursts are the most luminous sources of electromagnetic radiation known.
Since their progenitors are massive stars, they provide a tracer of star formation and star-forming galaxies over the whole of
cosmic history.
Their bright power-law afterglows provide ideal backlights for absorption studies of the interstellar and intergalactic
medium back to the reionization era.
The proposed THESEUS mission is designed to detect large samples of GRBs at $z>6$ in the 2030s, at a time when 
supporting observations with major next generation facilities will be possible, thus enabling a range of transformative science.
 THESEUS will allow us to explore the faint end of the luminosity function of galaxies and the star formation rate density to high redshifts; constrain the progress of re-ionisation 
 beyond $z\gtrsim6$; study in detail  early chemical enrichment from stellar explosions, including signatures of Population III stars; 
 and potentially characterize the dark energy equation of state at the highest redshifts.
\end{abstract}

\section{Introduction}
\label{intro}

A major goal of contemporary astrophysics and cosmology is to achieve a detailed understanding of the formation of the first collapsed objects (Pop-III and early Pop-II stars, black holes and the primordial galaxies in which they were  born) during the first billion years in the life of the Universe. 
The growth of these early luminous sources 
is intimately connected to the reionization of the intergalactic medium (IGM) and build-up of heavy chemical elements \cite{Ferrara15}.
Cosmic chemical evolution is very poorly constrained at high redshift, and even in the JWST era, metallicity estimates at $z>6$ will rely on crude, or model dependent, emission line diagnostics for only a limited number of the brightest galaxies (typically, $M_{\rm UV} < -19$ mag). Regarding reionization, measurement of the Thomson-scattering optical depth to the microwave background by the Planck satellite now suggests that it substantially occurred in the redshift range $z \sim 7 $– 8.5 \cite{Planck20}, but the timeline and the nature of the driving sources remain contentious \cite{Wise19,Naidu20}.

The timeline of reionization has been constrained chiefly from two directions. The first entails observations of the Gunn-Peterson trough \cite{GunnPeterson65} and  estimates of the redshift evolution of the Lyman $\alpha$  (henceforth Ly~$\alpha$) optical depths in the spectra of distant quasars, which indicate that reionization was nearly complete at $z\sim6$. Due to its sensitivity, the Gunn-Peterson test, which shows total absorption of  Ly~$\alpha$ photons,  therefore saturates for $z>6.1$ quasars (see, e.g., \cite{Becker01,Fan06,McGreer14}; also Section~\ref{sec:7} for supporting evidence from GRB afterglows). 
A second approach has proven successful at probing the neutral fraction of the IGM at higher redshifts, namely searches for Ly~$\alpha$ emission in spectroscopic observations of Lyman-break selected galaxies. These studies yield a drop in the fraction of Ly~$\alpha$ emitting galaxies at $z>6$ (e.g. \cite{Fontana10,Pentericci11,Vanzella11,Caruana12,Ono12,Bunker13,Treu13,Caruana14,Pentericci14,Schenker14,Tilvi14,Hoag19,Mason19,Jung20}), although there exist exceptions such as in observations of UV-faint galaxies behind galaxy cluster lenses \cite{Fuller20}, massive galaxies \cite{Endsley21}, and possibly in cases of ionised bubbles around bright galaxies \cite{Jung19}. There is also some evidence for a late reionization scenario (e.g.~\cite{Kusakabe20}), with the process possibly still ongoing as late as $z\sim5.5$ 
\cite{Gangolli21,Kulkarni19,Becker19}. 
Statistical measurements of the fluctuations in the redshifted 21 cm line of neutral hydrogen by future experiments, notably SKA-Low, are expected ultimately to provide further constraints on the time history of reionization \cite{Mondal20}.

The central question, however, remains whether it was predominantly radiation from massive stars that both brought about and sustained this phase change, as is conventionally
thought \cite{Eide2020}, or whether more exotic mechanisms must be sought. Even in
the JWST/ELT era, the numerous galaxies populating the faint-end of the luminosity function ($M_{\rm UV} \gtrsim -16$ mag) 
in the re-ionization epoch ($z>6$) will be very hard to access. Gauging star formation occurring in these faint galaxies, the average fraction of ionizing radiation that escapes from galaxies, together with the parallel build-up of metals 
and their spreading throughout the galactic ISM and the IGM by stellar and galactic winds, will still remain highly challenging problems. 

\begin{figure}
 \hspace{-2.5mm} \includegraphics[width=8.7cm]{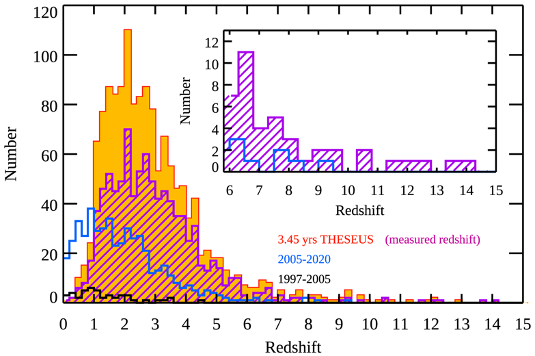}
\caption{A realisation of the expected redshift distribution of long-duration GRBs detected by THESEUS in 3.45 yr of scientific operations (orange and purple histograms) based on GRB population modeling \cite{Ghirlanda21}, compared with the redshift distribution obtained to date (blue and black histograms)}
\label{zdist}      
\end{figure}

GRBs offer a unique route to detecting the death-throes of individual massive stars
to very high redshifts 
(Figure~\ref{zdist}; \cite{Tanvir09,Salvaterra09,Cucchiara11,Tanvir18}), which in turn provides multiple powerful probes of early star formation, metal enrichment and galaxy evolution, potentially even before the main phase of reionization (e.g. \cite{Ghirlanda15}).
Indeed, they are detectable independently of the luminosity of their underlying hosts and so can pinpoint the presence of massive star formation in distant galaxies below the sensitivity limit of even the most powerful facilities foreseen in the long-term future. Additionally, their afterglow counterparts can be used as bright background lighthouses probing in absorption both the IGM and the interstellar medium (ISM) of faint galaxies otherwise non accessible with other observing techniques. 
The key to exploiting this new window on the early universe will be the detection and follow-up of samples of tens of GRBs in the era of reionization.
The proposed THESEUS mission \cite{Amati18} is expected to identify and locate between 40 and 50 GRBs at $z>6$ in 3.45 years of scientific operations \cite{Ghirlanda21}, and allow on-board determination of photometric redshifts more accurate  than 10\% thanks to the identification of the Lyman break feature being shifted to the imaging sensitivity range of its Infrared telescope (IRT, see \cite{Amati21}).  
In this white paper we outline the range of high redshift science that will be enabled by this mission.

\section{Global star formation rate from GRB rate as a function of redshift}
\label{sec:1}

Long-duration GRBs are produced by massive stars, and so track star formation, and in particular the populations of UV-bright stars responsible for the bulk of ionizing radiation production. This makes them important probes of global star formation, and significant changes in stellar populations, such as variations in  the initial mass function (IMF).

However, there is growing observational evidence
that the situation is more complicated in the low-$z$ universe. 
Population studies based on well selected, complete sub-samples of long GRBs, have reported that their rate increases more rapidly with $z$ and peaks at higher redshift
compared to cosmic star formation (e.g. \cite{Jakobsson12,Salvaterra12,Pescalli16}).
Moreover, analyses of the properties of galaxies
hosting low-$z$ GRB events have shown that they are typically less luminous, less massive and more metal poor than the normal field galaxy populations \cite{LeFloch06,Modjaz08,Graham13,Vergani15,Perley16,Palmerio19}. These findings are consistent with a scenario in which GRBs form preferentially from moderately metal-poor progenitors as indeed required by some theoretical models \cite{Woosley06,Yoon06}, and inefficiently in super-solar metallicity environments.
While the exact metallicity threshold at which GRB form
efficiently is still debated, ranging from $ 0.3\,Z_\odot $ to nearly solar \cite{Graham17}, there is general consensus that at high redshifts, where most of the galaxies are still metal-poor, GRBs can be used as fair tracers of the cosmic star formation (e.g. \cite{Campisi2011,Robertson12,Salvaterra13,Vergani17}). Recent hydrodynamical simulations accounting for complex chemical enrichment (based on \cite{Maio2010}) 
including self-consistent dust production support this view at $z \geq 4$ \cite{Graziani2020a}, and are consistent with the observed evolution of the dust-to-metals ratio in high-redshift galaxies probed by GRB-DLAs \cite{Wiseman2017}.
Under this assumption, their detected rate at $z>6$ can be used to estimate the cosmic star formation in an independent way, provided that the proportionality factor between stars and GRB event is calibrated. 
Even more directly, the evolution of the GRB rate with $z$ should be closely related to the decline of the cosmic star formation rate with increasing redshift, after the cosmic noon. 

\begin{figure}
 \centerline{\includegraphics[width=85mm]{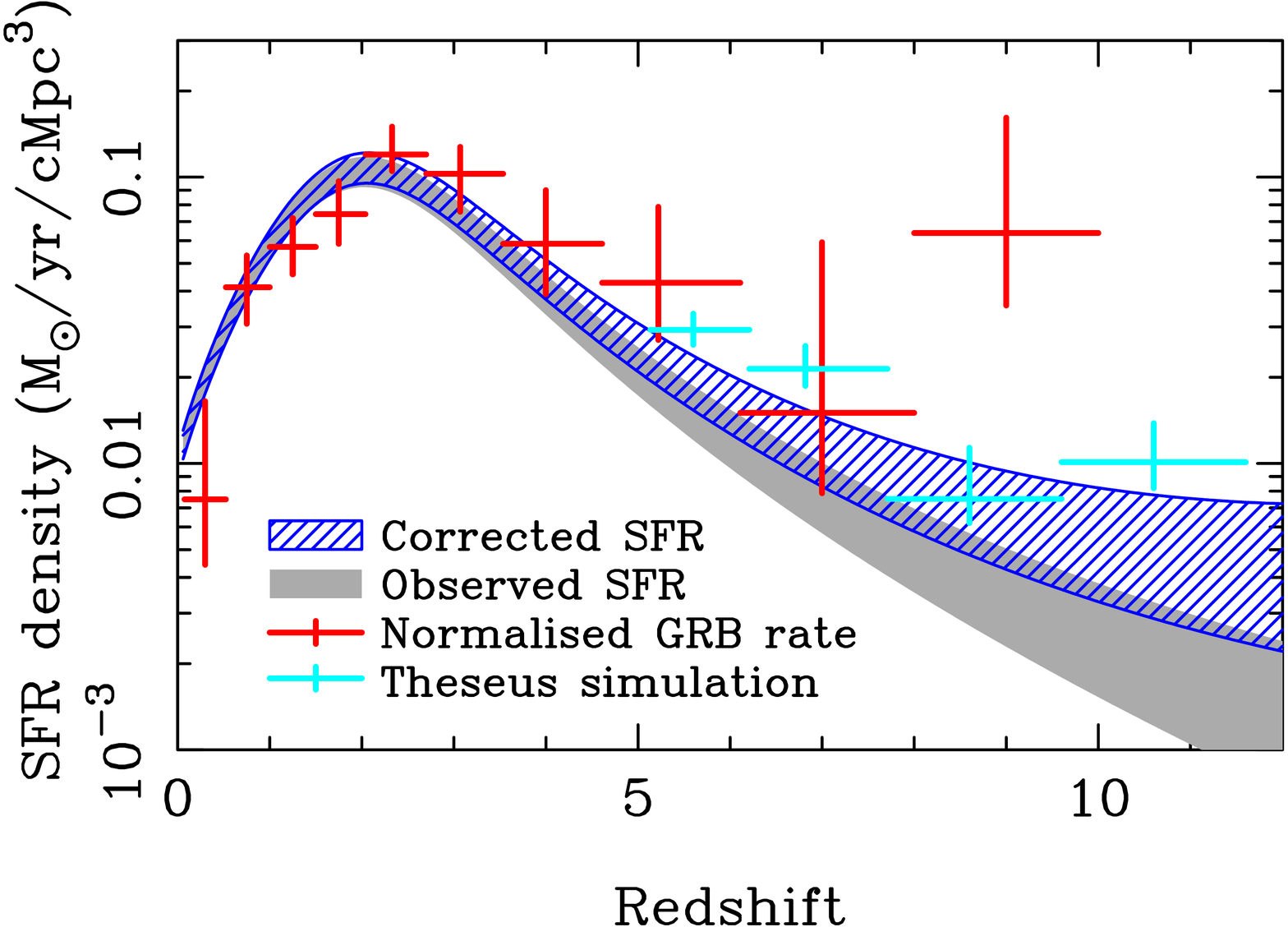}}
\caption{Comoving Star Formation Rate density as a function of redshift as derived from rest-frame UV surveys (shaded curve based on the observed galaxy population \cite{Madau17}, blue hatched region accounts for galaxies below the detection limit) and from GRBs based on different assumptions (GRB rate to SFR ratio, GRB progenitor metallicity \cite{Finkelstein19}). Red points show current constraints from the available GRB sample, which is very small at $z>6$ \cite{Perley16}, but shows some tension compared to estimates based on galaxy counts.  Cyan points show corresponding estimates expected from a representative THESEUS sample where the GRB rate has been enhanced at high redshift to illustrate a situation in which the current tension is maintained.}
\label{sfrz}      
\end{figure}

Intriguingly, analyses of this sort have consistently indicated a higher star formation rate (SFR) density at redshifts $z > 6$ than traditionally inferred from UV rest-frame galaxy studies (\cite{Perley16,Kistler09}; see Figure~\ref{sfrz}), which rely on counting star-forming galaxies and attempting to account for galaxies below the detection threshold as well as the fraction of UV light missed because of dust obscuration. Although this discrepancy has been alleviated by the growing realisation of the extremely steep faint-end slope of the galaxy Luminosity Function (LF) at $z > 6$ (e.g. \cite{Bouwens15}, it still appears that this steep slope must continue to very faint magnitudes (e.g. $M_{\rm AB} > -11$), in order to provide consistency with GRB counts and indeed to achieve reionization (something that can only be quantified via a full census of the GRB population).
Interestingly, some numerical models of galaxy formation performed at different cosmic scales, and then capable of resolving smaller galaxies clustering around Milky Way-like galaxies, support a similar view indicating that significant SFR is still hidden in undetected faint star forming systems clustering around over-dense ones at high redshift (see \cite{Graziani2020b} Fig. 1). 
Alternative, and equally interesting, possibilities are that the stellar IMF becomes more top-heavy at early times (GRB progenitors being drawn from populations with birth masses  
$\sim25-40$ M$_{\odot}$), 
 suggestive of very low metallicity populations \cite{Chon21}, or that there are multiple binary pathways  to producing long-GRB progenitors with different metallicity dependence \cite{Chrimes20}.

However, up to now, these studies have been
severely limited by the very small size of the high-$z$ GRB samples
\cite{Kistler09,Robertson12}. 
THESEUS will establish the GRB redshift distribution, $N(z)$, much more reliably at $z > 5.5$ than previous missions, thanks to significantly larger numbers detected by the SXI and also more uniform selection via rapid on-board redshift estimates with the IRT, combined in many cases with a refinement of the redshift through follow-up with other facilities. Thus, accounting for the observational selection function, we will obtain the evolution of the global star formation rate density at
redshifts where uncertainties from traditional galaxy observations begin to rise. This will firmly establish whether the tension between GRB counts and extrapolated galaxy counts remains, demanding a more radical reappraisal of high-$z$ star formation (Figure~\ref{sfrz}).

\begin{figure}
 \centerline{\includegraphics[width=78mm]{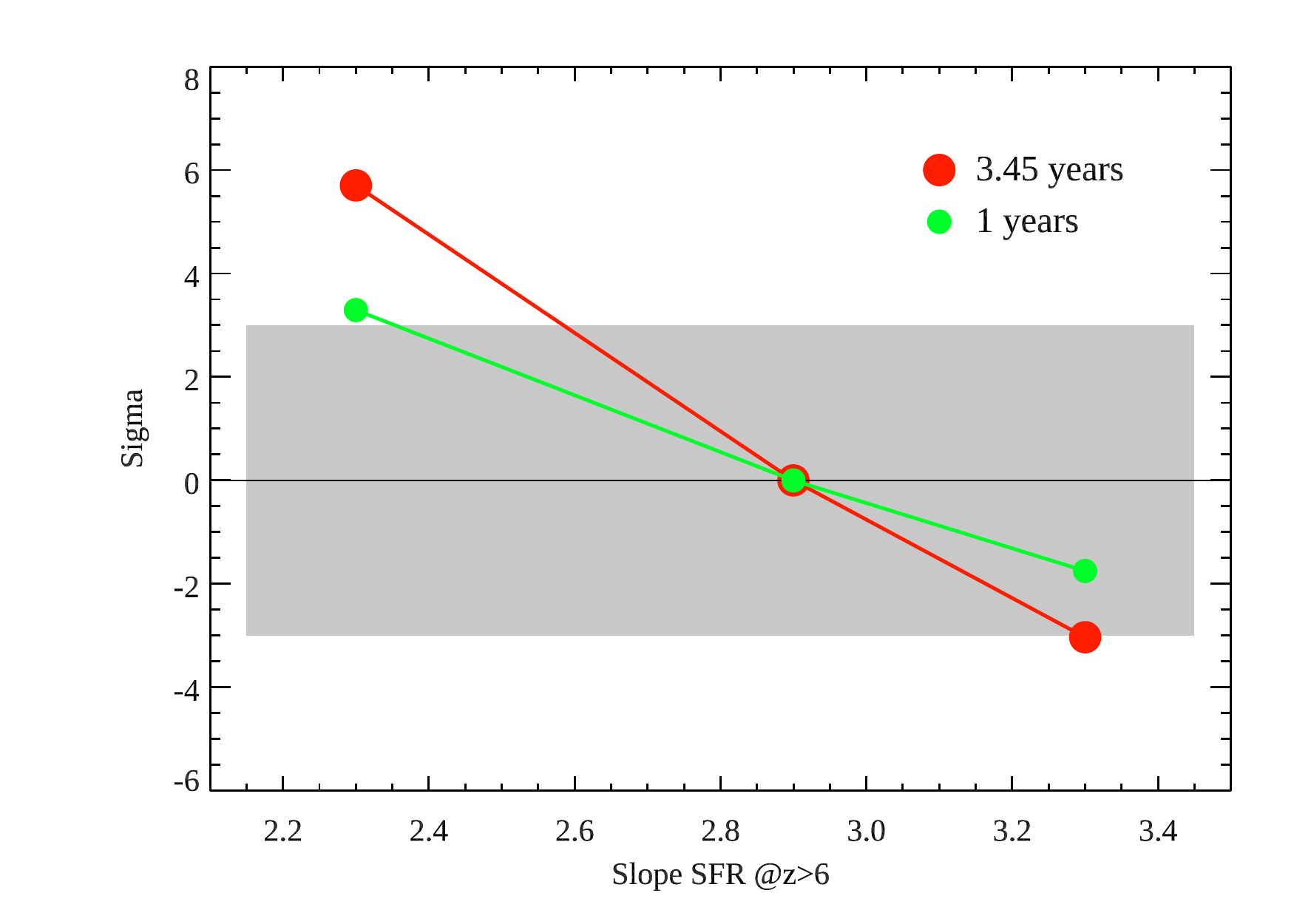}}
\caption{Accuracy in determining the decline slope of the SFR density as
measured by the number of GRBs detected by Theseus/SXI at $z>6$. Green (red) line and points represent 1 year (3.45 years) constraints. The grey shaded area reports the 3$\sigma$ confidence level: so in this example the actual slope is set to 2.9, and the first year results give a 3$\sigma$ range of about 2.4 to in excess of 3.4, whereas the nominal mission would constrain the result to between about 2.6 and 3.3. (Note, the SFR density slope measured by \cite{Madau14} is $\sim 2.7$.)}
\label{sfrslope}      
\end{figure}

As another way to demonstrate the power of the THESEUS sample,
in Figure~\ref{sfrslope} we show the accuracy at which  will measure the decline slope of the star formation rate density with redshift (specifically, $-d{\rm log(SFRD)}/d{\rm log}(1+z)$)
for the cosmic star formation at $z>6$ after 1 and 3.45 years of operations. Already after the first year, the measured number of high-$z$ GRBs will allow us to discard a slope as shallow as 2.4 at
3$\sigma$ confidence level, while discarding slopes as steep as 3.3 will
require the full 3.45-year sample. A slope as steep as 2.7 \cite{Madau14} will be measured with an error of 0.15 (1$\sigma$) while slightly tighter (looser) constraints are expected for shallower (steeper) declines.

Thus THESEUS provides an independent way, complementary to
deep-field galaxy surveys, to measure this important parameter at very high-$z$. Even more interestingly, the identification of any discrepancy between GRB-based measurements and those obtained by different methods could provide important insights about role of faint galaxies, missed even in future deep surveys, in shaping the cosmic star formation history, about the possible evolution of the stellar IMF in the very high-$z$ Universe, and about the existence of an important population of bright X-ray transients, such as those expected from Pop-III stars
(Sect.~\ref{sec:8}).

\section{The galaxy luminosity function: detecting undetectable galaxies}
\label{sec:2}

The intrinsically very small galaxies, which appear to increasingly dominate star formation at $z > 6$, are very hard to detect directly. GRBs circumvent this difficulty, sign-posting the existence and redshifts of their hosts, no matter how faint.

\begin{figure}
\hspace{-3mm} \includegraphics[width=0.79\textwidth]{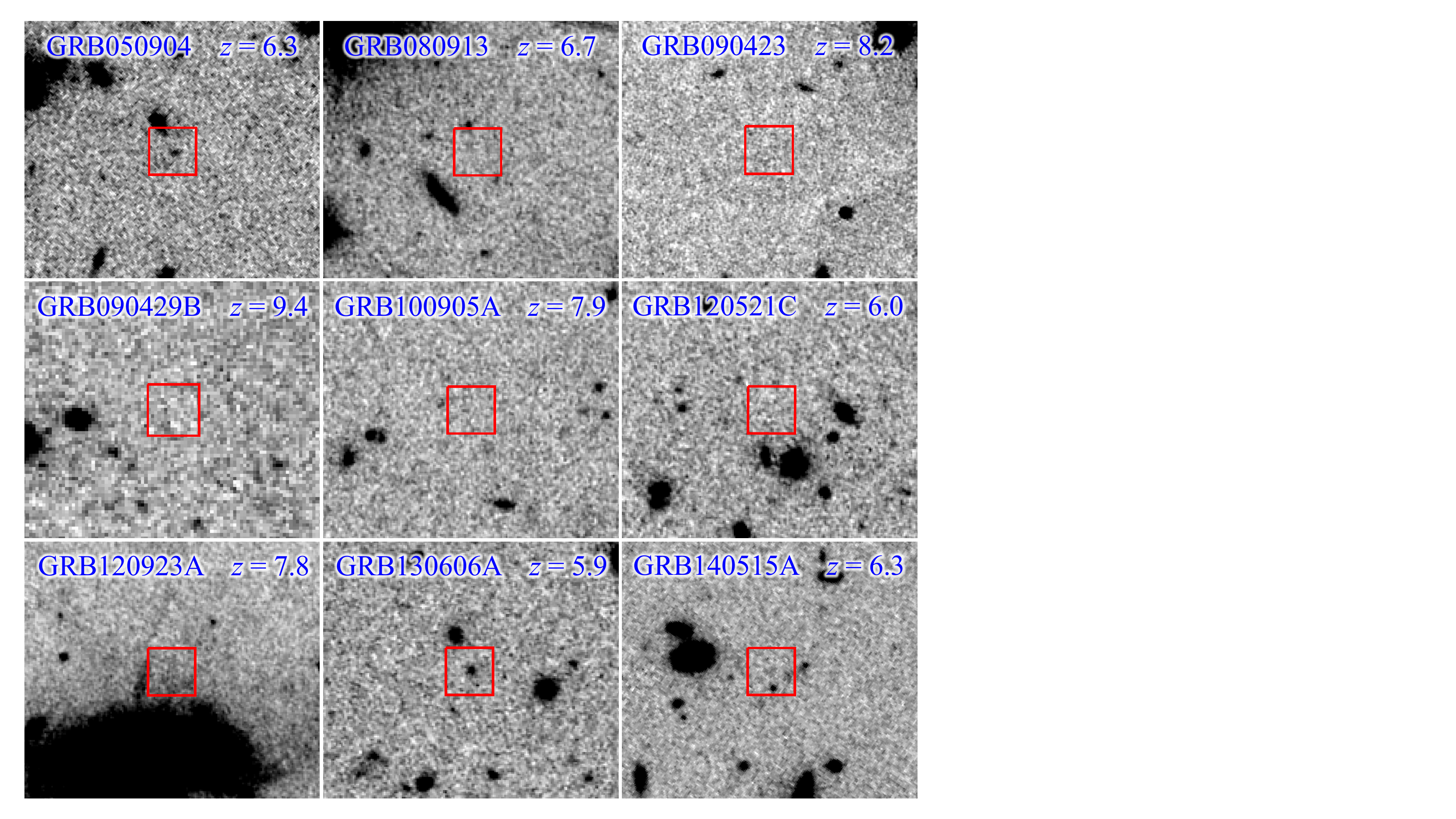}
\caption{Mosaic of deep HST imaging of the locations of known GRBs at $z > 6$, obtained when the afterglows had faded (red boxes are centered on the GRB locations, and are 2 arcsec on a side). Only in 2--3 cases is the host galaxy detected, confirming that the bulk of high-$z$ star formation is occurring in galaxies below current limits. This approach allows us to quantify the contribution of the faint end of the galaxy luminosity function to the star formation budget, even in the absence of direct detections.}
\label{mosaic}       
\end{figure}

The faint end of the galaxy luminosity function is a key issue for our understanding of reionization since, to the depth achieved in the Hubble Ultra-deep Field (HUDF), it appears that the faint-end of the LF steepens significantly with redshift approaching a power-law of slope $\alpha \sim 2$ at $z > 6$ (i.e. where the number of galaxies per unit luminosity $\phi(L)\propto L^{-\alpha}$ for faint galaxies; e.g. \cite{Bouwens15}). 
Studies, which take advantage of gravitational lensing by galaxy clusters in the Hubble Frontier Fields, can reach even fainter magnitudes, albeit probing increasingly smaller volumes in the higher magnification regime, and are also consistent with a steep faint-end slope \cite{Atek15,Bouwens17,Vanzella21}.
The value of the total luminosity integral depends sensitively on the choice of low-luminosity cut-off (and indeed the assumption of continued power-law form for the LF). By conducting deep searches for the hosts of GRBs at high-$z$ we can directly quantify the ratio of star-formation occurring in detectable and undetectable galaxies, with the sole assumption that GRB-rate is proportional to star-formation rate (Figure~\ref{mosaic}). Although currently limited by small-number statistics, early application of this technique has confirmed that the majority of star formation at $z > 6$ occurred in galaxies below the effective detection limit of HST \cite{Tanvir12,McGuire16} with expected magnitudes fainter than $m_{\rm AB} \sim 30$, at the limit of what is reachable with JWST and the ELTs. Since the exact position and redshift of the galaxy is known from the GRB afterglow, follow-up observations to measure the host UV continuum are much more efficient than equivalent deep field searches for Lyman-break
galaxies. If the luminosity function is modelled as a Schechter function with a sharp faint-end cut-off, then this analysis allows us to constrain that cut-off magnitude, even though the galaxies are too faint to be observed. This technique applied to a sample of $\sim$\,$40$ GRBs detected by THESEUS at $z > 6$ will yield much tighter constraints on that cut-off magnitude than obtained with GRBs so far (Figure~\ref{reionization}, left).

\section{The build-up of metals, molecules and dust}
\label{sec:3}

Bright GRB afterglows, with their intrinsic power-law spectra, provide ideal backlights for measuring not only the hydrogen column, but also obtaining exquisite abundances and gas kinematics, probing into the hearts of their host galaxies \cite{Hartoog15}. Thus, they can be used to map cosmic metal enrichment and chemical evolution to early times, and search for evidence of the nucleosynthetic products of even earlier generations of stars.

\begin{figure*}
  \includegraphics[width=0.49\textwidth]{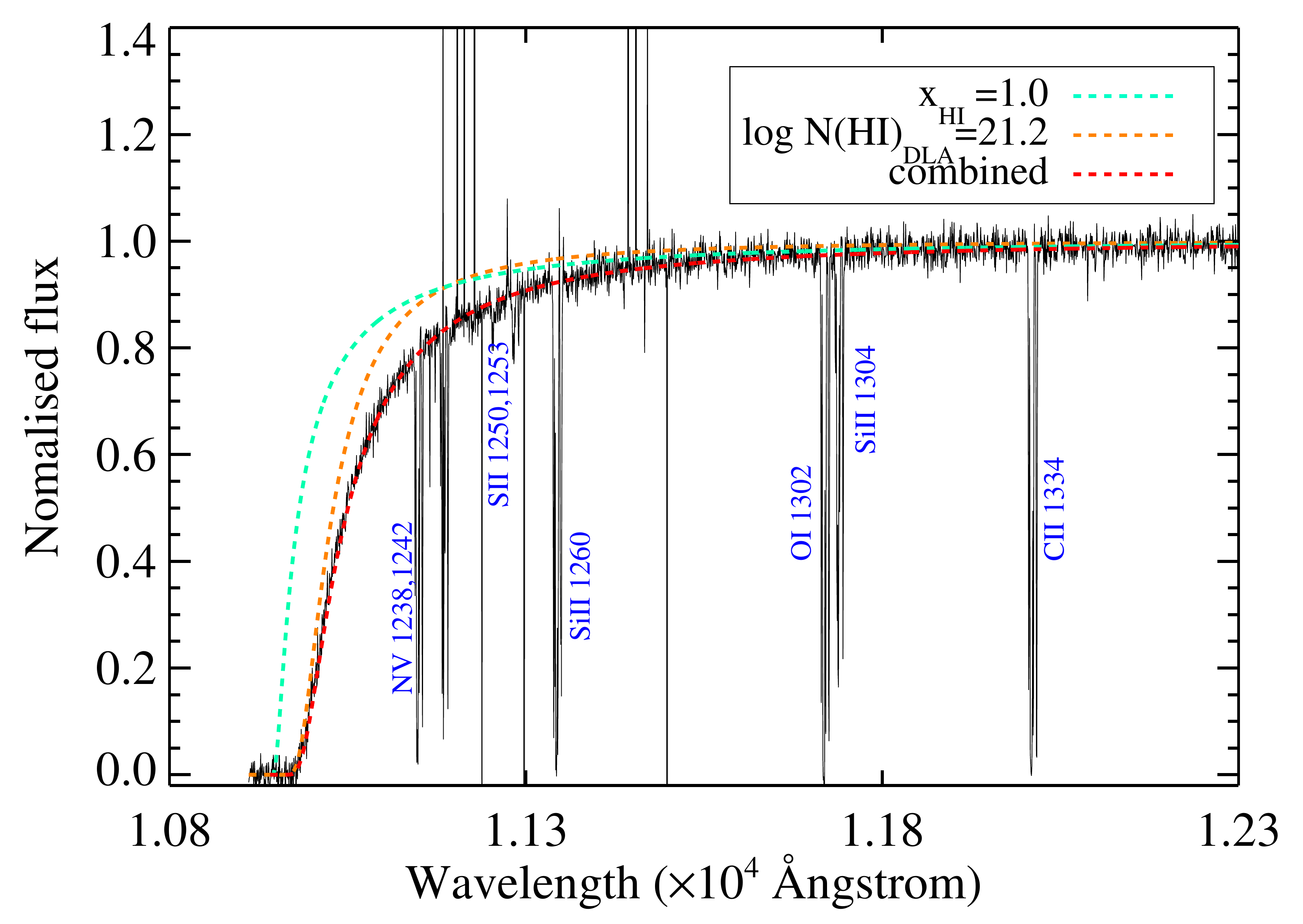}
  \includegraphics[width=0.5\textwidth]{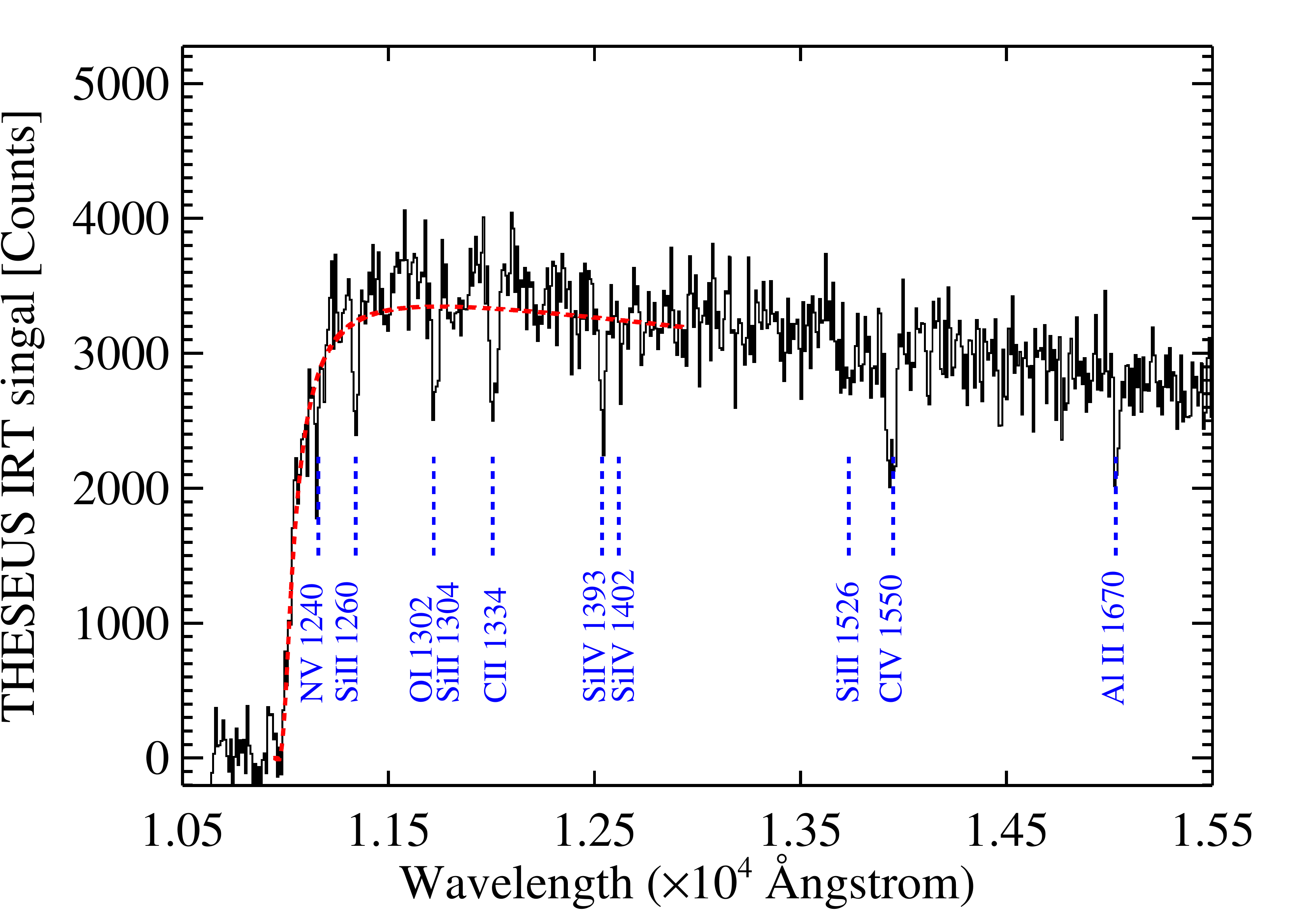}
\caption{Left: Simulated ELT 30-minute spectrum of a $z=8.0$ GRB afterglow with $J({\rm AB})=20$ (typical after $\sim0.5$ day). The S/N provides exquisite abundance determinations from metal absorption lines (in this example, 1\% solar metallicity), while fitting the Ly~$\alpha$ damping wing simultaneously fixes the IGM neutral fraction and the host {\sc Hi} column density, as illustrated by the two overlaid models, a pure 100\% neutral IGM (green) and a ${\rm log}(N_{\rm HI}/{\rm cm}^{-2})=21.2$ host absorption with a fully ionized IGM (orange). A well-fitting combined model is shown in red. Right: The same afterglow with a magnitude of 16 in a simulated IRT spectrum with a realistic spectral observing sequence at a total integration time of 1800 s, illustrating that the most prominent metal lines are also clearly detected.}
\label{specsims}       
\end{figure*}

Based on the specifications of the IRT spectrograph with its spectral resolution $R\sim400$, we simulate a set of possible afterglows with varying host galaxy hydrogen column densities and metallicities, and analyse the precision that we will be able to achieve as a function of the signal-to-noise ratio of the spectra (Figure~\ref{specsims}, right hand panel). 
First, follow-up of the brightest afterglows ($H_{\mathrm{AB}} \lesssim17.5$ mag) with the IRT spectroscopic mode  will provide constraints within $\sim0.2$ dex on the hydrogen column density along the GRB line of sight in the host galaxy. The identification of associated metal absorption lines will also enable spectroscopic redshift determinations to $<1$\% precision, which will help refining the IRT photo-$z$ estimates and distinguishing cleanly between the GRBs at $z > 6$ and contaminants from dusty afterglows at lower redshifts. 
Even though IRT spectra will be able to detect the usual rest-frame UV metal absorption lines \cite{Fynbo09} for afterglows with metallicities above $\sim$1\% solar for an afterglow detected at $H_{\mathrm{AB}}\lesssim 16.5$ mag, accurate abundance, metallicity and abundance pattern determination necessitates follow-up deeper, higher-spectral resolution data from space- or ground-based facilities.

Taking advantage of the availability of 30 m class ground-based telescopes in the 2030s superb abundance determinations will be possible through simultaneous measurement of metal absorption lines and modelling the red-wing of Ly-$\alpha$ to determine host HI column density, potentially even many days post-burst (Figure~\ref{specsims}, left panel), 
thanks to the effects of cosmological time dilation
for the highest redshift events.  This may be supported by ATHENA observations to quantify the high ionization gas content (e.g. \cite{Heintz18}) and potentially radio for other atomic fine structure and molecular lines (e.g. \cite{Urata}).

\begin{figure}
\includegraphics[width=8.2cm]{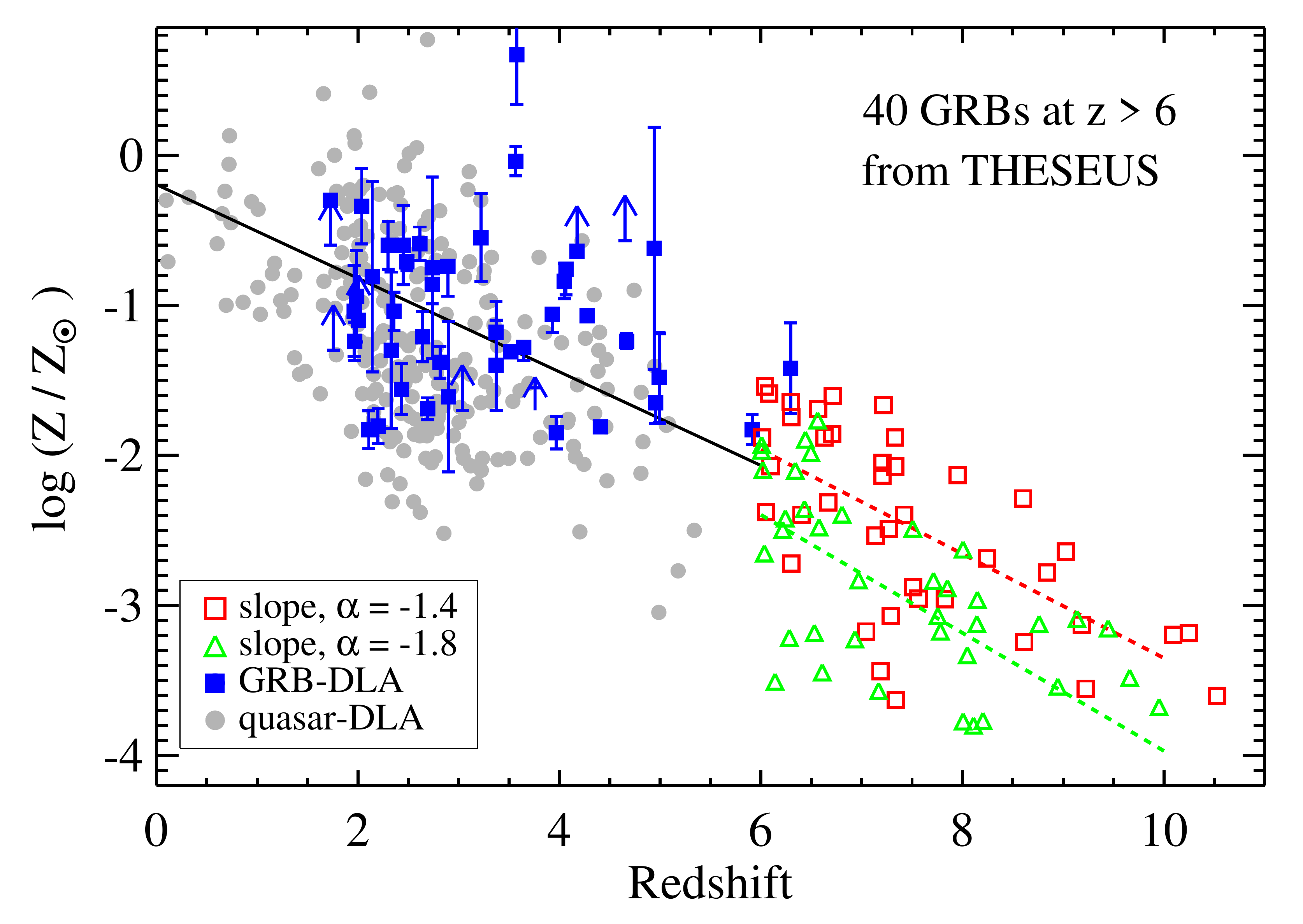}
\caption{Absorption-line based metallicities relative to solar, corrected for dust depletion as a function of redshift for Quasar Damped Ly~$\alpha$ absorbers (DLAs, grey symbols) and GRB-DLAs (blue symbols) (adapted from \cite{DeCia18}, \cite{Bolmer18}). Open square symbols show representative expectations for THESEUS, assuming continued evolution of the mass-metallicity relationship, and a dominant population of low mass galaxies at $z>6$ (green triangles and red squares assume faint-end slopes of $-1.8$ and $-1.4$ for the galaxy luminosity function, respectively). GRBs represent the unique way for probing evolution of ISM absorption–based metallicities in the first billion years of cosmic history.}
\label{Z}
\end{figure}

Using the sample of GRBs discovered by THESEUS to trace the ISM in galaxies at $z>6$ will be the only way to map in detail exact metallicities and abundance patterns across the whole range of star forming galaxies in the early Universe, including those at the very faint end of the LF (Figure~\ref{Z}). 
Thus, for example, we will be able to search directly for evidence of alpha-enhancement, hinted at by modelling of emission line spectra at intermediate redshifts \cite{Runco21}, and evidence of pop III chemical enrichment (see Section~\ref{sec:8}).

From the observed LF for galaxies at $z>6$ \cite{Bouwens15}, we can estimate the distributions of metallicities for GRBs at the highest redshifts by assuming that GRBs trace the galaxy luminosities and a relation between the stellar mass and gas-phase metallicity is valid. While this mass-metallicity relation has been established at $z<3$ for galaxies in emission \cite{Maiolino08} and based on absorption line metallicities \cite{Moller13}, a considerable uncertainty remains on the validity at higher redshifts, and we have to rely on an extrapolation that will be tested with future analyses of galaxies with the JWST.
Since the low-end of the LF at the higest redshifts is debated, we vary the slope of the LF faint end, which impacts the resulting metallicity distributions. With a significant sample of GRB afterglow metallicities at $z>6$ THESEUS will provide an independent test of the slope of the LF at $z>6$. Most importantly, this test is entirely independent of being able to detect the host galaxy in emission. 

There is a rising theoretical interest in both the cosmic early chemical enrichment and the properties of the ISM of high redshift galaxies, also supported by recent observations of dusty galaxies in the epoch of reionization \cite{Laporte2018,Tamura2019}, and complemented by recent results from the ALPINE survey \cite{Faisst2020}, down to $z\sim4$. 
ISM constraints provided by THESEUS will be critical to further improve both cosmological models investigating the early assembly of dusty galaxies \cite{McKinnon2017,Aoyama2018,Graziani2020a} and the detailed evolution of the simulated ISM only accessible in zoom-in simulations of isolated galaxies \cite{Aoyama2017,Granato2020}. Many of the processes regulating dust evolution in the galactic environment are in fact strictly depending on the multi-phase ISM in which both metal atoms and dust grains co-evolve.

The imprint of the dust in the host can be seen in the (rest-frame UV/optical) broad-band spectral energy distributions of GRB afterglows. Such studies have found a variety of dust laws, many reasonably approximated by the three canonical local laws (SMC, LMC, Milky Way; e.g. \cite{Schady12,Zafar18}), but along some sight-lines the extinction is unusual and harder to explain \cite{Fynbo14}. Thus, GRBs offer a remarkable route to assessing the dust content of even low mass galaxies in the early Universe \cite{Zafar11}. 
At moderate redshifts, molecular hydrogen can be detected through electronic transitions
(Lyman and Werner bands). Due to its specific formation and destruction processes, H$_2$
provides an excellent probe of the cold neutral medium (CNM) in galaxies \cite{Friis15}. Being inclined to Jeans
instability, this phase plays a crucial role in the star-formation process, but is not detectable
in emission. In addition, the detection of various rotational levels of H$_2$, together
with fine-structure levels of companion species (as {\sc Ci}) provides very sensitive tools to determine
the prevailing physical conditions (in particular density, temperature,  and UV field) \cite{Noterdaeme17,Balashev19}.

Although this becomes unfeasible at high redshift due to the Gunn-Peterson trough obscuring the Lyman-Werner bands, other absorption tracers, notably vibrationally excited H$_2^*$, and neutral carbon, {\sc Ci}, provide reliable indicators of high molecular hydrogen columns \cite{Noterdaeme18,Heintz19}.

We emphasize that using the THESEUS on-board NIR spectroscopy capabilities will provide, in addition to arcsec accurate location, the redshift estimates and luminosity measurements that are essential to optimising the time-critical follow-up observations using highly expensive next-generation facilities, allowing us to select the highest priority targets and deploy the most appropriate telescope and instrument.

Finally, we note that also in the X-ray band, SXI could allow for minimal studies of the medium surrounding the GRB site. As observed in GRB050904, X-ray data holds the potential to reveal a decrease in the X-ray absorbing column density and to constrain the metallicity \cite{Campana07}. SXI could follow GRB050904-like events up to $\sim 3000$ s during which the full photoionisation of the medium will occur. 

\section{The Lyman-continuum escape fraction}
\label{sec:4}

GRB afterglow spectroscopy allows us to measure the column density of neutral hydrogen in the host galaxy, thus providing a powerful probe of the opacity of the interstellar medium to EUV photons.

\begin{figure}
 \hspace{-3mm}   \includegraphics[width=8.8cm]{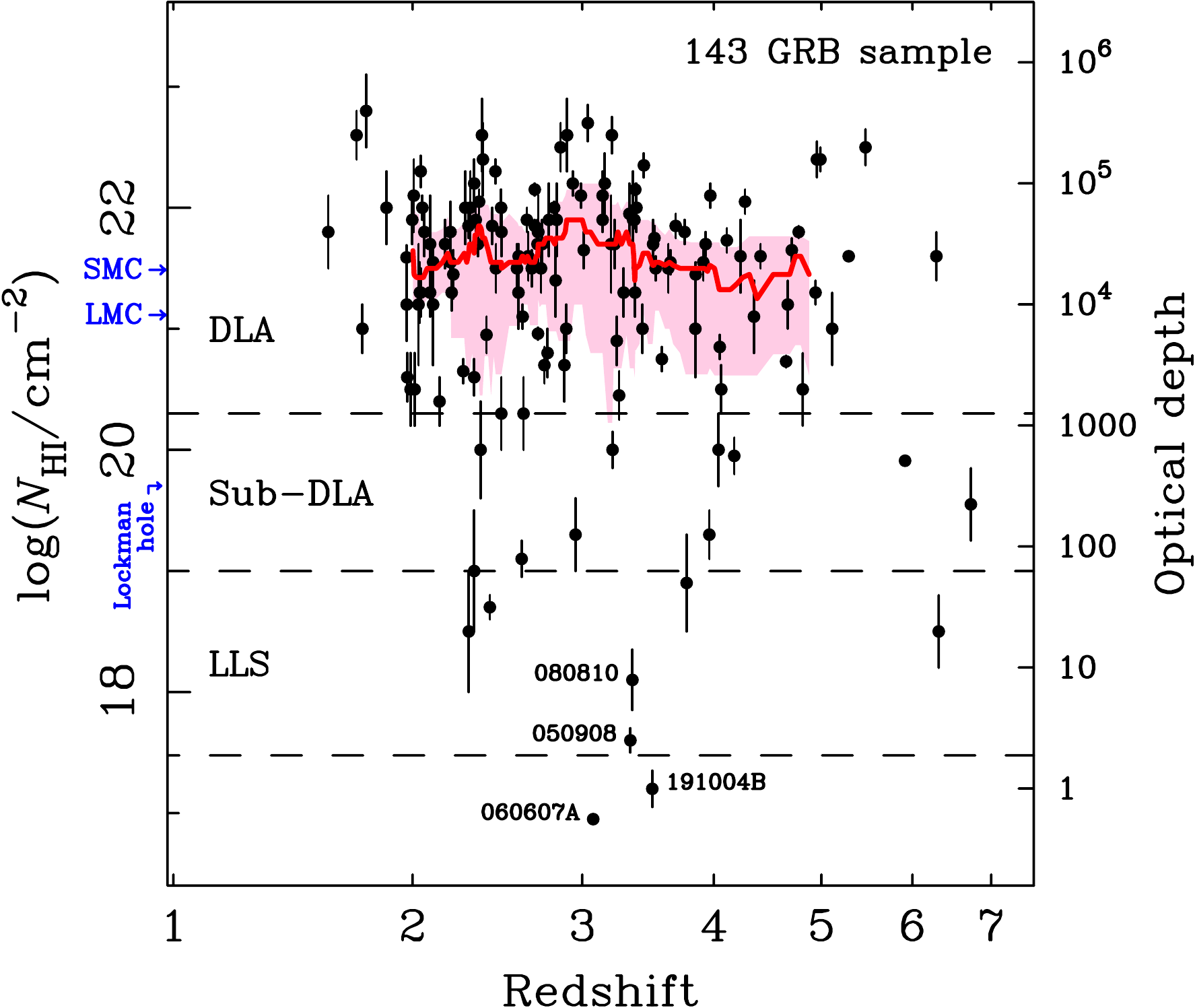}
\caption{The host neutral hydrogen column densities measured from the Ly\,$\alpha$ absorption line in the afterglows of 143 GRBs spanning a wide range of redshift. Typically, the columns imply a high optical depth to ionizing Ly-continuum radiation, and thus a low overall escape fraction ($f_{\rm esc}<2$\%).
Strikingly, there is little evolution in the running median value (red line) or inter-quartile range (pink region) over the redshift for which the distribution is well sampled, but there is a hint of a possible downturn at the highest redshifts.
THESEUS will enable us to extend this plot with good statistics at $z > 5$ providing a direct estimate of the escape fraction on the lines of sight to massive stars in this era.}
\label{fesc}      
\end{figure}

A key issue for the reionization budget is quantifying the fraction of ionizing radiation that escapes the galaxies in which it is produced. Even in the JWST/ELT era, direct observations of escaping Lyman continuum radiation at $z > 5$ will remain unfeasible for the low mass galaxies responsible for the bulk of star formation in the era of reionization. However, GRB afterglow spectroscopy allows us to measure the column density of neutral hydrogen in the host galaxy, thus providing a strong lower limit to the opacity of the interstellar medium to FUV photons (further EUV attenuation due to dust can also be estimated by modelling the afterglow SEDs). A statistical sample of afterglows can be used to infer the average escape fraction over many lines of sight, specifically to the locations of massive stars dominating global ionizing radiation production. 
Useful constraints have so far only been possible at $z = 2$--5, indicating a strong upper limit of $f_{\rm esc}< 2$\% (Figure~\ref{fesc};  \cite{Tanvir19,Vielfaure20}), although note this is lower than found in the recent study of stacked LBG galaxy spectra at $z\sim3$ by \cite{Steidel18}. However, future observations of the population of $z > 5$ GRBs detected by THESEUS, with both on-board spectroscopy and with 30 m class ground-based telescope follow-up observations, will provide much more precise constraints on the fraction of ionizing radiation that escaped galaxies during the epoch of reionization. 

\section{Did stars reionize the Universe?}
\label{sec:5}

The evolution of the IGM from a completely neutral to a fully ionized state is intimately linked to early structure formation, and thus a central issue for cosmology. Answering the key question of whether this phase change was primarily brought about stars hinges on two subsidiary issues: how much massive star formation was occurring as a function of redshift, and, on average, what proportion of the ionizing radiation produced by these massive stars escaped from the immediate environs of their host galaxies? Both will be addressed through THESEUS GRB observations.

The former problem, quantifying massive star formation as a function of redshift, can be extrapolated based on observed candidate $z > 7$ galaxies found in HST deep fields, but two very significant uncertainties are, firstly, the completeness and cleanness of the photometric redshift samples at $z > 7$, and, secondly, the poorly constrained faint-end behaviour of the galaxy luminosity function (at stellar masses $\lesssim 10^8$ M$_{\odot}$), especially since galaxies below the HST (and potentially even the JWST) detection limit very likely dominate the star-formation budget.
Even though some constraints on fainter galaxies can be obtained through observations of lensing clusters \cite{Atek15}, which will be improved further by JWST, simulations suggest that considerable star formation was likely occurring in fainter systems still \cite{Liu16}. 

As discussed in Section~\ref{sec:4}, the second problem, that of the Lyman-continuum escape fraction, is even more difficult since it cannot be determined directly at these redshifts, and studies at lower redshifts have found conflicting results. Recent stacked spectroscopic analyses have suggested escape fractions as high as 10\% \cite{Steidel18}, which could be sufficient to drive reionization \cite{Finkelstein19}, but it is unclear whether the samples of galaxies studied, at $z\sim3$, are representative of all star- forming galaxies, and in particular of the typical, intrinsically fainter galaxies at
$z>7$. As seen in Figure~\ref{fesc}, GRB studies find that sight-lines to massive stars are generally highly opaque to ionizing radiation, at least up to $z\sim5$.

\begin{figure*}
\centerline{
 \begin{minipage}{50mm}
\includegraphics[width=45mm]{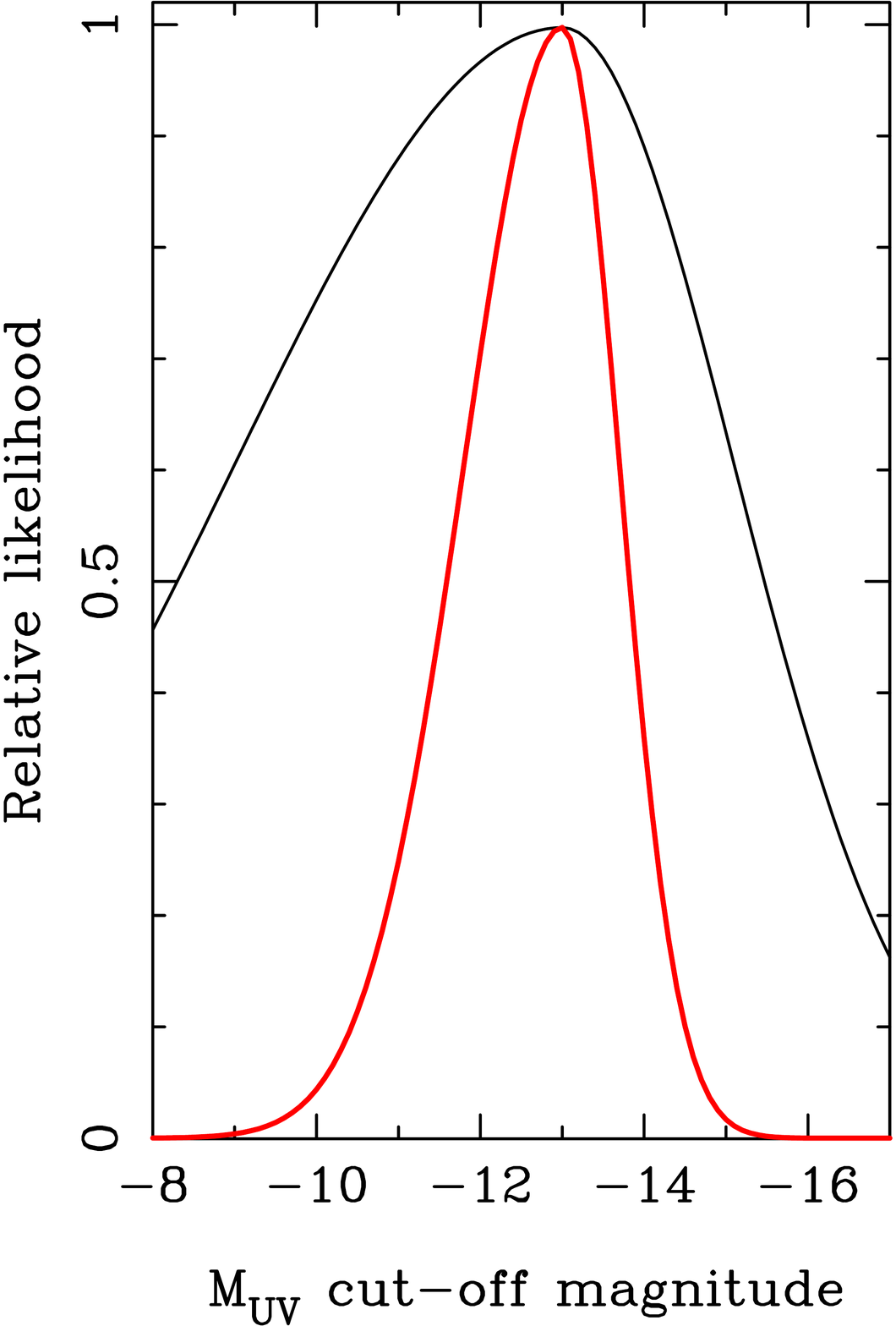}\end{minipage}
   \begin{minipage}{115mm}
   \includegraphics[width=110mm]{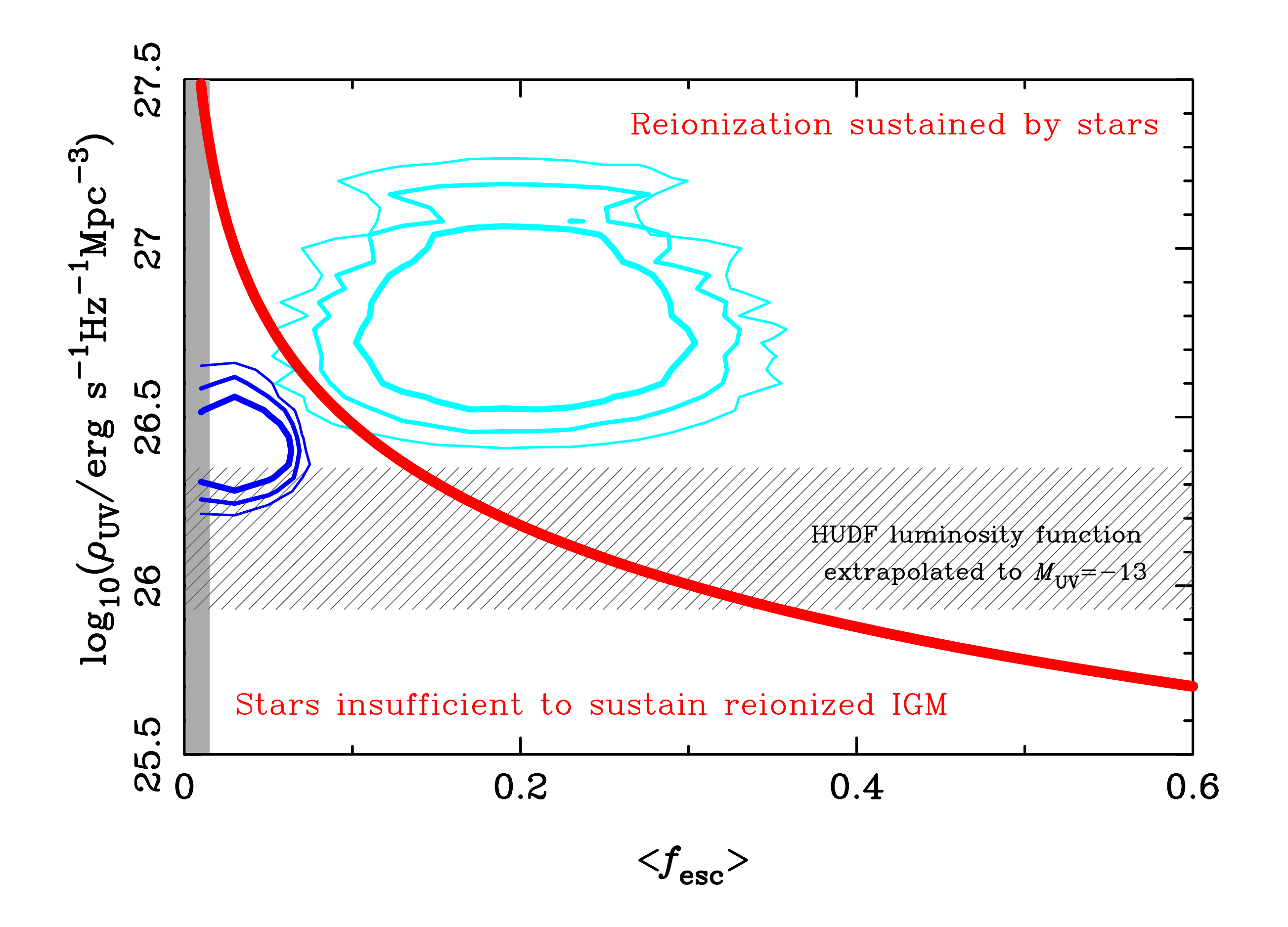}\end{minipage}
}
\caption{Left: constraints on the faint end cut-off of the galaxy luminosity function (assuming a Schechter function with abrupt cut-off and bright end constrained by galaxy observations). Black curve shows current constraints based on nine GRBs at $z \ge 6$; red curve shows a simulation after THESEUS, assuming deep searches for a sample of $\sim45$ hosts from 30 m class telescopes. Right: the UV luminosity density from stars at $z\sim8$ and average escape fraction $\langle f_{\rm esc}\rangle$ are insufficient to sustain reionization \cite{Robertson13} unless the galaxy luminosity function steepens to magnitudes fainter than $M_{UV}=-13$ (hatched region), and/or $\langle f_{\rm esc}\rangle$ is much higher than that found from GRB studies at $z\sim2$--5 (shaded region). Even in the late 2020s, $\langle f_{\rm esc}\rangle$ at $z>6$ will be largely unconstrained by direct observations. The contours show the 2-$\sigma$ expectations for samples of 20, 30 and 50 GRBs (thin to thick contours) at $z\sim7$--9 for which deep spectroscopy provides the host neutral column and deep imaging constrains the fraction of star formation occurring in hosts below ELT limits. 
The cyan contours  illustrate a model with high escape fraction ($f_{\rm esc}\approx20$\%) and SFRD, whereas the blue contours are for a model with lower values of both ($f_{\rm esc}\approx4$\%).
}
\label{reionization}       
\end{figure*}

The improvements in both the census of star formation and the escape fraction from THESEUS GRB studies will provide a strong test of the hypothesis that reionization was brought about by star light. Our detailed simulations indicate that THESEUS is expected to detect between 40 and 80 GRBs at $z > 6$ over a four-year mission, with between 10 and 25 of these at $z > 8$ (and several at $z > 10$) \cite{Ghirlanda21}. The on-board follow-up capability will mean that redshifts are estimated for almost all of these, and powerful next generation ground- and space-based telescopes available in this era will lead to extremely deep host searches and high-S/N afterglow spectroscopy for many (e.g. using ELT, ATHENA etc.). To illustrate the potential of such a sample, we simulate in Figure~\ref{reionization}   (right) the precision in constraining the product of the UV luminosity density and average escape fraction, $\rho_{\rm UV} f_{\rm esc}$, that would be obtained with samples of 20, 30 and 50 GRBs at $7<z<9$ having high-S/N afterglow spectroscopy and ($\sim$3hr) ELT depth host searches (for definiteness the $\rho_{\rm UV}$ axis corresponds to $z=8$). This will allow us to confidently distinguish between conventional models in which starlight brings about reionization, and models that fail (e.g. if the escape fraction remains as low as we find at lower redshifts).

\section{Topology and timeline of reionization}
\label{sec:6}

In practice, it is expected that reionization should proceed in a patchy way, for example, ionized bubbles may grow initially around the highest density peaks where the first galaxies form, expanding and ultimately filling the whole IGM. The topology of the growing network of ionized regions reflects the character of
the early structure formation and the ionizing radiation field. 
The Ly~$\alpha$ scattering cross section is composed of a principal, exponential-profile component arising from thermal motion, and a tail component due to natural broadening. This latter damping wing absorption is weaker than the principal absorption component by several orders of magnitude, and since it follows a power-law profile, it ventures into longer wavelengths than the principal component. Effectively, this damping wing absorption is far less sensitive than Gunn-Peterson absorption, and can therefore be used to probe higher neutral fractions where the Gunn-Peterson test saturates. Via this damping wing absorption, therefore, we may constrain the neutral fraction of the IGM for values upward of 10\%. 

To date, this has been challenging both because of the sparsity of events, and the S/N ratio that is required, such that the experiment has been limited to a very small number of GRBs (e.g.~GRB~050904 \cite{Totani06} and GRB~130606A \cite{Chornock13,Hartoog15}). 
With high-S/N afterglow spectroscopy, the red damping wing of the hydrogen Ly~$\alpha$ line can be decomposed into contributions due to the host galaxy and the IGM,  with the latter providing a measure of the progress of reionization local to the burst.

\begin{figure}
\centerline{  \includegraphics[width=60mm]{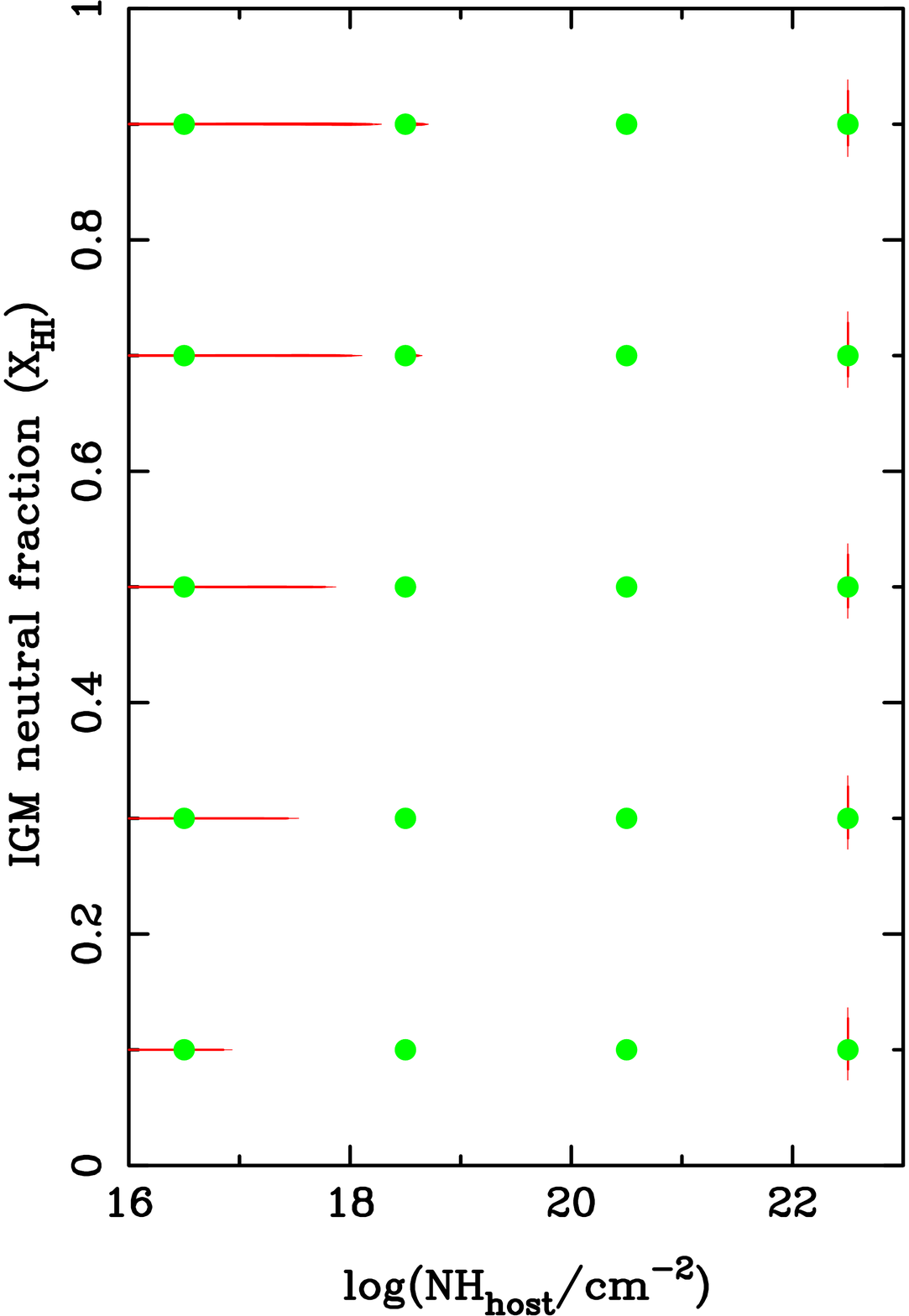}}
\caption{Results of simulations showing how well the IGM neutral fraction $\chi_{\rm HI}$ and host {\sc Hi} column density can be retrieved from typical ELT-class afterglow spectra (cf. Figure~\ref{specsims}). The input values are shown as green points, and the resultant fits by the red contours. In nearly all cases the IGM neutral fraction is well recovered (within the symbol size), as are host columns for all but the lowest cases, although even here results should be sufficient to indicate a partially transparent sight line. The variance between sight-lines at the same redshift of IGM neutral fraction is expected to be up to $\delta\chi_{\rm HI}\sim0.3$ at mid reionization \cite{McQuinn08}, depending on the reionization model, which can be determined with the THESEUS sample.}
\label{decomp}      
\end{figure}

GRBs themselves do not ionise the surrounding IGM, but previously occurring star formation in the host galaxy as well as contributions from nearby galaxies will form {\sc Hii} regions. A patchy reionization scenario, which sees galaxies carving out {\sc Hii} bubbles which later merge, would imply that measurements will vary across a sample of GRBs. By carrying out follow-up high-resolution spectroscopy of a sample of several tens of GRBs at $z > 6$--9, we can begin to investigate statistically the redshift-dependence of the average and variance of the reionization process \cite{McQuinn08}. To illustrate the capability of follow-up spectroscopy to quantify both host {\sc Hi} column, and IGM neutral fraction, Figure~\ref{decomp} shows the results of simulating an afterglow with the characteristics of GRB 090423 (cf. Figure~\ref{specsims}) with a range of different input parameters, for which the measured parameters over a large number of realisations are shown (green points). This demonstrates that the IGM neutral fraction can be recovered from such spectra, even in the presence of a fairly high host column, and similarly the host column can generally be well characterised except when very low.  While the situation may be complicated further due to the additional effect of any local ionized bubble in which the host resides, but again this can be modelled with sufficient S/N spectra \cite{McQuinn08}, as is likely to be obtained for many afterglows with 30\,m telescope spectroscopy.

Using GRBs for this endeavour offers advantages over reionization studies that employ Ly~$\alpha$ emitters (LAEs; e.g. \cite{fuller20}) for two main reasons: firstly, GRBs suffer less from selection bias compared to LAEs; secondly, one can often get an accurate redshift measurement via the presence of absorption lines in the afterglow spectrum. This may be contrasted to LAEs, and also Lyman Break Galaxies (LBGs), where redshift determination can sometimes be ambiguous when it depends solely on Ly~$\alpha$ emission itself (in the former) and the presence of the break (in the latter), with potential confusion with low-$z$ interlopers.

Observations of the damping wing in GRB afterglow spectra also offer advantages over the same type of exercise in quasars (cf. \cite{mortlock11}). In the first instance, one expects that GRBs do not represent such biased  sources as quasars; in this respect, with GRBs one would not be probing “special” regions, e.g. as in higher-density environments where quasars are regularly located. Moreover, on account of GRBs being short-lived events, one does not expect their emitted radiation to have a large impact on the ionisation state of the surrounding IGM, which means that one would obtain a more representative picture than with quasars.
Finally, quasars often exhibit complicated continuum emission and bright emission lines such as Ly~$\alpha$, making the damping wing modelling and the analysis of the foreground Lyman forest particularly challenging. By contrast, the featureless continuum of GRB afterglows makes this exercise much more simple and leads to {\sc Hi} fraction and column density estimates with much smaller uncertainties (see also Figure~\ref{specsims}).

\section{Population III stars and primordial galaxies}
\label{sec:7}

High redshift GRBs provide several routes to exploring the earliest populations of metal-free and ultra-low metallicity stars, from direct spectroscopic determinations of ISM abundances, to changes in the numbers and properties of the bursts themselves, reflective of the increase in average and maximum masses expected for Population III stars \cite{Chon21}.

In the current $\rm \Lambda$CDM paradigm, the very first stars (the so-called Population III stars) are expected to form from pristine gas, primarily at very high redshifts, $z\sim10$--30 \cite{Bromm11}. 
The absence of heavy elements and consequent inefficiency of cooling at these early cosmic times, is expected to lead to masses that largely exceed those of Pop-I and Pop-II stars (possibly reaching several hundreds of solar masses). When these first stars reach their final stage of evolution, their low-opacity envelope combined with limited mass loss from stellar winds may thus keep large amounts of gas bound until the final collapse, favouring the conditions for jet breakout and for the launch of a very energetic long GRB \cite{Meszaros10}. Such models  predict that the total equivalent isotropic energy released by such Pop-III star explosions could exceed by several orders of magnitude that of GRBs from Pop-I/II progenitors, possibly reaching $\sim10^{56} –10^{57}$ erg and making them detectable up to the highest possible redshifts \cite{Toma11}. 
Given the peculiar energetics and chemistry associated with Pop-III star GRBs, their observed properties should differ from GRBs at lower redshift. In particular, their prompt emission could extend over much longer timescales, potentially lasting up to a month \cite{Yoon15}.

Similarly, the energy released by the jetted explosion could imply much longer times to dissipate. This could give rise to a luminous afterglow emission peaking much later and at higher fluxes than observed for Pop-II GRBs, which in particular would be extremely bright at radio wavelengths if happening in a high-density ISM
\cite{Burlon16}.
A luminous thermal component could also be produced if the jet deposits a large fraction of its total energy in the stellar envelope of the GRB progenitor.

Spectroscopy of such afterglows with 30 m-class telescopes (or JWST, if still operational) may reveal ultra-low metallicity if their line of sight intersects pristine gas in their host galaxy, and these signatures would represent a direct piece of evidence for the association between a GRB and a Pop-III star progenitor. Similarly, gas cloud pockets enriched by Pop-III star explosions and highlighted by Pop-II GRBs might provide us with another way to explore the metal abundance patterns characterizing the ISM enriched only by the very first generation of massive stars  \cite{MB2014,Ma15}.
Such abundance patterns may in fact reveal if the first heavy elements were produced by typical core-collapse explosions or by other mechanisms such as pair-instability supernovae, hence constraining the Initial Mass Function up to the earliest cosmic epochs \cite{Wang12,Ma17}.

\begin{figure}
\centerline{  \includegraphics[width=8.4cm]{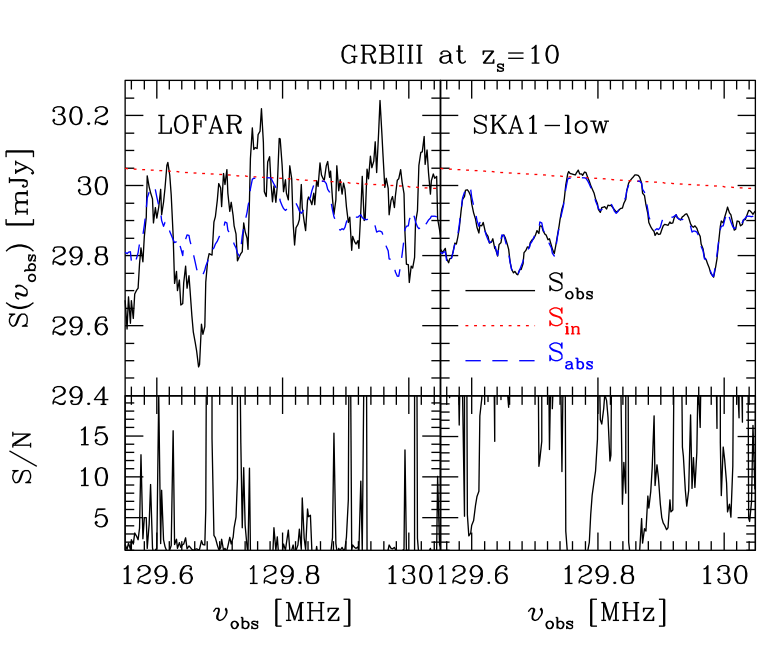}}
\caption{Top: spectrum of a GRB with a Pop-III star progenitor located at $z_s=10$ with a flux density of 30~mJy. The lines refer to the intrinsic spectrum of the source (red dotted), the simulated spectrum for 21~cm absorption (blue dashed), and the absorption spectrum as it would be seen with a bandwidth of 10~kHz and an integration time of 1000~h (black solid). The left and right panels refer to observations with the LOFAR and SKA-1 telescopes. Bottom: corresponding S/N ratio. See \cite{Ciardi15} for further details.}
\label{21cmforest}      
\end{figure}

It has been suggested that 21~cm absorption features (the 21~cm forest) in the spectra of high-$z$ radio sources could be used to gather information on the neutral hydrogen content in the IGM along the line of sight, similarly to what is presently done with the Ly~$\alpha$ forest at lower redshift. Although this is an intriguing possibility, the lack of high-$z$ sufficiently bright radio-loud sources makes it a purely theoretical one. If a GRB with a Pop-III star progenitor were detected, this could be an excellent candidate for follow-up searches of absorption features in its afterglow spectrum with the present or next generation of radio instruments such as SKA \cite{Ciardi15}. While the example in Figure~\ref{21cmforest}, refers to a source located at redshift 10, similar arguments apply to GRBs from Pop-III stars throughout the epoch of reionization. 

Predicting how many of such events THESEUS will identify throughout the duration of its science operations is particularly challenging since the probe of such first light sources still remains a fully uncharted territory.
Albeit rare and quickly transitioning to a more regular Pop-II regime \cite{Maio2010} \cite{Maio2011}, Pop-III star formation at high redshift 
may represent a significant proportion of the massive stars that end their lives as long GRBs.
Relying on Swift calibrations and numerical simulations of primordial structure formation, the fraction of Pop-III GRBs (born in  metal-poor sites with metallicity $Z < 10^{-4}\, Z_\odot$) may increase from low to high redshift up to 10\% (40\%) at $z > 6$ ($z > 10$) \cite{Campisi2011}, in line with the trend in Fig.~\ref{Z}.
Considering that THESEUS will observe around 40-80 GRB candidates at $z > 6$, there is an estimated probability of detecting at least 4-8 GRBs from Pop-III origin at $z > 6$. Such detections have the potential to be revolutionary, being the first primordial explosions ever observed.

In the era of 30 m class optical/nIR ground-based telescopes, THESEUS could be the only experiment enabling the discovery of these first very massive stellar explosions, hence providing exquisite targets for further follow-up with larger telescopes.
Pop-III stars likely played a major role in the growth of the very first bound structures at early cosmic times, through chemical feedback and metal enrichment of the primordial IGM, and they may have also contributed a head start to the cosmic re-ionization process. To date, no direct evidence of the connection between Pop-III stars and GRBs has been observationally established. The identification by THESEUS of even a single GRB with metal abundance unveiling a Pop-III star progenitor or a Pop-III star enriched medium would put fundamental constraints on the unknown properties of the first stars and represent a major breakthrough in our understanding of first-light sources.

\section{Probing the expansion history of the Universe and dark energy with GRBs }
\label{sec:8}

While the standardised candle correlations proposed for GRBs \cite{Amati02,Yonetoku04,Ghirlanda04b}
exhibit larger scatter than those for SNIa, and are subject to potentially complex selection effects (e.g. \cite{butler10}), GRBs can be observed to much higher redshift, and thus, in principle, provide much greater lever arm for testing cosmological world models.

By using the spectral peak energy – radiated energy (or luminosity) correlation $E_{\rm p,i} – E_{\rm iso}$ \cite{Amati02} (or $L_{\rm iso}$; \cite{Yonetoku04}), it has been argued that GRBs offer a promising tool to probe the expansion rate history of the Universe beyond the current limit of $z\sim2$ (Type-Ia SNe and Baryonic Acoustic Oscillations from quasar absorbers). With the present data set of GRBs, cosmological parameters consistent with the concordance cosmology can already be derived \cite{Ghirlanda04,Amati08}. 
The order of magnitude improvement provided by THESEUS on the size of the sample of GRBs above $z\approx3$, with more uniform selection (on-board measured redshifts in many cases and broad-band spectral parameters), will allow us to further refine the reliability of these methods and, possibly, to characterize the equation of state 
of Dark Energy.
For example, the THESEUS sample may help assess whether the dark energy equation of state evolves with redshift, 
such as in extended dynamical dark energy  models, like scalar field quintessence or alternative theories of gravity \cite{DiValentino20,Lusso19}.
Combined with the most recent constraints from the Cosmic Microwave Background, this will offer the unique opportunity to constrain the geometry, and therefore the mass-energy content of the Universe back to $z\sim5$ or further, extending beyond the investigations of EUCLID and of next-generation large-scale structure surveys to the entire cosmic history.

Furthermore, at higher redshift, effects from alternative scenarios for dark matter nature could be detectable. They should impact the first collapsing phases of cosmic gas and cascade over the entire star formation history.
In particular, warm dark matter (WDM) is an interesting possibility advocated to solve small-scale problems of the standard model based on cold dark matter (CDM).
It relies on the assumption that dark matter is made by warm particles having a non-negligible (although small) thermal velocity.
IGM based calibrations suggest a mass limit for WDM particles corresponding to about 3 keV which would produce a sharp decrease of power at relatively small scales.
With this constraint, there is little room to disentangle between WDM and CDM in terms of low-redshift star formation, but it is possible to test WDM-induced delays in the gas molecular cooling and the onset of star formation at early times.
In this respect, the high-redshift regime probed by THESEUS ($z > 6$) becomes an important new window that needs to be explored.
During primordial epochs, WDM models predict a neat cut-off in the SFR and GRB history, departing from CDM expectations \cite{MV2015}.
By tracing  early GRBs with THESEUS, it will be possible to probe the nature of dark matter (WDM vs. CDM), independently from other techniques, and the lack of observed objects during the epoch of reionization will help pose serious constraints on WDM.
This will be very interesting, as WDM implications on structure growth are thought to be more dramatic than the ones derived from  dark energy \cite{quintessence06}
or from non-linear high-order corrections to cosmological perturbation theory \cite{bulkflows11}.

Long GRBs detected by THESEUS in such early epochs could be useful also to constrain local non-Gaussianities in the matter distribution. Indeed, positive (negative) deviations from the Gaussian shape of the primordial density field are expected to enhance (inhibit) structure formation in early epochs. Assuming that GRBs are fair tracers of cosmic star formation, this means that positive non-Gaussianities might increase the expected GRB rate at high redshift \cite{Maio2012}.
Any excess of long GRB detections at $z > 10$ by THESEUS would impact the high-$z$ GRB rate and help constrain possible non-Gaussian scenarios. Even one long GRB with a rate of at least $10^{-6}\,\rm yr^{-1} sr^{-1}$ in such early epochs would be sufficient to put additional constraints on the matter distribution.
GRB detections in line with the cosmic star formation rate density, instead, would support the commonly adopted Gaussian assumption. In any case, once observed by THESEUS, long GRBs exploding during the epoch of reionization would provide new insights about the properties of dark matter.

\section{Conclusions}
\label{sec:9}
Thanks to its unprecedented sensitivity and follow-up performance compared to past and current GRB-dedicated experiments, THESEUS will identify more than 40 events at $z>6$, hence providing  a much larger sample of high redshift GRBs than achieved so far. This will open a unique window on the early Universe, supplementing the
picture of the first billion years of cosmic evolution to be painted in the coming decade by JWST, the ELTs and SKA. Spectroscopic follow-up of their afterglows will  address a number of key issues such as the history of chemical enrichment in the ISM, the timeline of reionization and the escape fraction of ionizing radiation in early star-forming galaxies. The occurrence rate of high-$z$ GRBs and the properties of their hosts will also further refine constraints on the cosmic history of star formation and the luminosity function at the very faint end, which may contribute solving the still open debate on the overall contribution of massive stars to reionization. THESEUS will finally open new routes for chasing explosion signatures from the still-elusive population III stars, hence offering exciting opportunities to probe primordial galaxies.

\bigskip\bigskip\bigskip\bigskip\bigskip


%
%


\begin{thebibliography}{}
%
%


%


\bibitem{Ferrara15} Ferrara, A., Metal Enrichment in the Reionization Epoch, in Understanding the Epoch of Cosmic Reionization, ASSL 423, Springer International Publishing Switzerland, 163 (2016)
\bibitem{Planck20} Planck Collaboration, Planck 2018 results. VI. Cosmological parameters, A\&A, 641, 6 (2020)
\bibitem{Wise19} Wise, J., Cosmic Reionization,  arXiv:1907.06653 (2019)
\bibitem{Naidu20} Naidu, R. P. et al., Rapid Reionization by the Oligarchs: The Case for Massive, UV-bright, Star-forming Galaxies with High Escape Fractions,  ApJ, 892, 109 (2020)
\bibitem{GunnPeterson65} Gunn, J.E. and Peterson B.A., On the Density of Neutral Hydrogen in Intergalactic Space, ApJ, 142, 1633 (1965)
\bibitem{Becker01} Becker, R. H. et al., AJ, Evidence for Reionization at $z\sim6$: Detection of a Gunn-Peterson Trough in a $z=6.28$ Quasar, 122, 2850 (2001)
\bibitem{Fan06} Fan, X. et al., Constraining the Evolution of the Ionizing Background and the Epoch of Reionization with $z\sim6$ quasars. II. A sample of 19 Quasars, AJ, 132, 117 (2006)
\bibitem{McGreer14} McGreer, I., et al., Close companions to two high-redshift quasars, AJ 148, 73 (2014) 
\bibitem{Fontana10} Fontana A. et al., The Lack of Intense Ly~$\alpha$ in Ultradeep Spectra of $z=7$ Candidates in GOODS-S: Imprint of Reionization?, ApJ, 725, L205 (2010)
\bibitem{Pentericci11} Pentericci, L. et al., Spectroscopic Confirmation of $z\sim7$ Lyman Break Galaxies: Probing the Earliest Galaxies and the Epoch of Reionization, ApJ, 743, 132 (2011)
\bibitem{Vanzella11} Vanzella, E. et al., Spectroscopic Confirmation of Two Lyman Break Galaxies at Redshift Beyond 7, ApJ, 730, L35 (2011)
\bibitem{Caruana12} Caruana, J. et al., No Evidence for Lyman~$\alpha$ Emission in Spectroscopy of $z>7$ Candidate Galaxies, MNRAS, 427, 3055 (2012)
\bibitem{Ono12} Ono, Y. et al., Spectroscopic Confirmation of Three $z$-dropout Galaxies at $z=6.844-7.213$: Demographics of Ly$\alpha$ emission in $z\sim7$ galaxies, ApJ, 744, 83 (2012)
\bibitem{Bunker13} Bunker, A.J. et al., VLT/XSHOOTER and Subaru/MOIRCS spectroscopy of HUDF.YD3: No Evidence for Lyman $\alpha$ Emission at $z=8.55$, MNRAS, 430, 3314 (2013)
\bibitem{Treu13} Treu, T. et al., The Changing Ly$\alpha$ Optical Depth in the Range $6<z<9$ from the MOSFIRE Spectroscopy of $Y$-dropouts, ApJL, 775, L29 (2013)
\bibitem{Caruana14} Caruana, J. et al., Spectroscopy of $z\sim7$ Candidate Galaxies: Using Lyman $\alpha$ to Constrain the Neutral Fraction of Hydrogen in the High-Redshift Universe, MNRAS, 443, 2831 (2014)
\bibitem{Pentericci14} Pentericci, L. et al., New Observations of $z\sim7$ Galaxies: Evidence for a Patchy Reionization, ApJ, 793, 113 (2014)
\bibitem{Schenker14} Schenker, M. et al., Line-emitting Galaxies beyond a Redshift of 7: An Improved Method for Estimating the Evolving Neutrality of the Intergalactic Medium, ApJ, 795, 20 (2014)
\bibitem{Tilvi14} Tilvi, V. et al., Rapid Decline of Ly$\alpha$ Emission toward the Reionization Era, ApJ, 794, 5 (2014)
\bibitem{Hoag19} Hoag, A., et al., Constraining the Neutral Fraction of hydrogen in the IGM at Redshift 7.5, ApJ, 878, 12 (2019)
\bibitem{Mason19} Mason, C. et al., Inferences on the Timeline of Reionization at $z\sim8$ from the KMOS Lens-Amplified Spectroscopic Survey, MNRAS, 485, 3947 (2019)
\bibitem{Jung20} Jung, I. et al., Texas Spectroscopic Search for Ly$\alpha$ Emission at the End of Reionization. III. The Ly$\alpha$ Equivalent-Width Distribution and Ionized Structures at $z>7$, ApJ, 904, 144 (2020)
\bibitem{Fuller20} Fuller, S. et al., Spectroscopically Confirmed Ly$\alpha$ Emitters from Redshift 5 to 7 behind 10 Galaxy Cluster Lenses, ApJ, 896, 156 (2020)
\bibitem{Endsley21} Endsley, R. et al., MMT Spectroscopy of Lyman-alpha at $z\simeq7$: Evidence for Accelerated Reionization around Massive Galaxies, MNRAS, 502, 6044 (2021)
\bibitem{Jung19} Jung, I. et al., Texas Spectroscopic Search for Ly$\alpha$ Emission at the End of Reionization. II. The Deepest Near-Infrared Spectroscopic Observation at $z\gtrsim7$, ApJ, 877, 146 (2019)
\bibitem{Kusakabe20} Kusakabe, H. et al., The MUSE Hubble Ultra Deep Field Survey. XIV. Evolution of the Ly$\alpha$ emitter fraction from $z=3$ to $z=6$, A\&A, 638, A12 (2020)
\bibitem{Gangolli21} Gangolli et al., Constraining reionization in progress at $z=5.7$ with Lyman-$\alpha$ emitters: voids, peaks, and cosmic variance, MNRAS, 501, 5294G (2021)
\bibitem{Kulkarni19} Kulkarni, G., et al., Large Ly$\alpha$ opacity fluctuations and low CMB $\tau$ in models of late reionization with large islands of neutral hydrogen extending to z$<$5.5, MNRAS 485, L24 (2019)
\bibitem{Becker19} Becker, G., et al., The Evolution of O I over 3.2$<$z$<$6.5: Reionization of the Circumgalactic Medium, ApJ 883, 163 (2019)
\bibitem{Mondal20} Mondal, R. et al., Predictions for measuring the 21-cm multifrequency angular power spectrum using SKA-Low, MNRAS, 494, 4043 (2020)
\bibitem{Eide2020}  Eide, Marius B. et al., Large-scale simulations of H and He reionization and heating driven by stars and more energetic sources, MNRAS, 498, Issue 4, pp.6083-6099 (2020)
\bibitem{Tanvir09} Tanvir, N. R. et al., A $\gamma$-ray burst at a redshift of $z\sim8.2$, Nature, 461, 1254 (2009)
\bibitem{Salvaterra09} Salvaterra, R. et al., GRB090423 at a redshift of $z\sim8.1$, Nature, 461, 1258 (2009)
\bibitem{Cucchiara11} Cucchiara, A. et al., A Photometric Redshift of $z \sim 9.4$ for GRB 090429B, ApJ, 736, 7 (2011)
\bibitem{Tanvir18} Tanvir, N. R. et al., The Properties of GRB 120923A at a Spectroscopic Redshift of $z \approx 7.8$, ApJ, 865, 107 (2018)
\bibitem{Ghirlanda15} Ghirlanda, G. et al., Accessing the population of high-redshift Gamma Ray Bursts,  MNRAS, 448, 2514 (2015)
\bibitem{Amati18} Amati, L. et al., The THESEUS space mission concept: science case, design and expected performances,  AdSpR, 62, 191 (2018)
\bibitem{Ghirlanda21} Ghirlanda, G. et al., GRB science with THESEUS, this Volume
\bibitem{Amati21} Amati, L. et al., THESEUS: science case,  requirements and mission concept, this Volume 
\bibitem{Jakobsson12} Jakobsson, P. et al., The Optically Unbiased GRB Host (TOUGH) Survey. III. Redshift Distribution, ApJ, 752, 62 (2012)
\bibitem{Salvaterra12} Salvaterra, R. et al., A Complete Sample of Bright Swift Long Gamma-Ray Bursts. I. Sample Presentation, Luminosity Function and Evolution, ApJ, 749, 68 (2012)
\bibitem{Pescalli16} Pescalli, A. et al., The rate and luminosity function of long gamma ray bursts, A\&A, 587, 40 (2016)
\bibitem{LeFloch06} Le Floc'h, E., et al., Probing cosmic star formation with long gamma-ray bursts: new constraints from the Spitzer Space Telescope, ApJ 642, 636 (2006)
\bibitem{Modjaz08} Modjaz, M., et al., Measured metallicities at the sites of nearby broad-lined Type IC Supernovae and implications for the SN-GRB connection, AJ 135, 1136 (2008)
\bibitem{Graham13} Graham, J. \& Fruchter A., The metal aversion of Long-duration Gamma-ray Bursts, ApJ 774, 119 (2013)
\bibitem{Vergani15} Vergani, S. et al., Are long gamma-ray bursts biased tracers of star formation? Clues from the host galaxies of the Swift/BAT6 complete sample of LGRBs . I. Stellar mass at $z < 1$, A\&A, 581, 102 (2015)
\bibitem {Perley16} Perley, D. et al., The Swift GRB Host Galaxy Legacy Survey. II. Rest-frame Near-IR Luminosity Distribution and Evidence for a Near-solar Metallicity Threshold, ApJ,  817,  8 (2016)
\bibitem{Palmerio19} Palmerio, J. et al., Are long gamma-ray bursts biased tracers of star formation? Clues from the host galaxies of the Swift/BAT6 complete sample of bright LGRBs. III. Stellar masses, star formation rates, and metallicities at $z > 1$, A\&A, 623, 26 (2019)
\bibitem{Woosley06} Woosley, S. E. \& Heger, A., The Progenitor Stars of Gamma-Ray Bursts, ApJ, 637, 914 (2006)
\bibitem{Yoon06}  Yoon, S.-C. et al., Single star progenitors of long gamma-ray bursts. I. Model grids and redshift dependent GRB rate, A\&A, 460, 199 (2006)
\bibitem{Graham17} Graham, J. \& Fruchter, A., The Relative Rate of LGRB Formation as a Function of Metallicity, ApJ, 834, 170 (2017)
\bibitem{Campisi2011} Campisi, M.~A., Maio, U., Salvaterra, R., Ciardi, B. \ Population III stars and the long gamma-ray burst rate.\ MNRAS, 416, 2760–2767 (2011)
\bibitem{Robertson12} Robertson, B. \& Ellis, R., Connecting the Gamma Ray Burst Rate and the Cosmic Star Formation History: Implications for Reionization and Galaxy Evolution,  ApJ, 744, 95 (2012)
\bibitem {Salvaterra13} Salvaterra, R. et al., Simulating high-$z$ gamma-ray burst host galaxies, MNRAS, 429, 2718 (2013)
\bibitem{Vergani17} Vergani, S. et al., The chemical enrichment of long gamma-ray bursts nurseries up to $z = 2$, A\&A, 599, 120 (2017)
\bibitem{Maio2010} Maio, U. et al. \ The transition from population III to population II-I star formation.\ MNRAS, 407, 1003 (2010)
\bibitem{Graziani2020a} Graziani, L. et al., The assembly of dusty galaxies at z $>$ 4: statistical properties, MNRAS, 494, 1071 (2020)
\bibitem{Wiseman2017} Wiseman, P. et al., Evolution of the dust-to-metals ratio in high-redshift galaxies probed by GRB-DLAs  , A\&A, 599,  id.A24, 23 pp. (2017)
\bibitem {Kistler09} Kistler, M. et al., The Star Formation Rate in the Reionization Era as Indicated by Gamma-Ray Bursts, ApJ, 705, 104 (2009)
\bibitem {Bouwens15} Bouwens, R. et al., UV Luminosity Functions at Redshifts $z \sim 4$ to $z \sim 10$: 10,000 Galaxies from HST Legacy Fields, ApJ,   803,   34 (2015)
\bibitem {Madau17} Madau, P. et al., Radiation Backgrounds at Cosmic Dawn: X-Rays from Compact Binaries, ApJ, 840, 39 (2017)
\bibitem {Finkelstein19} Finkelstein, S. et al., Conditions for Reionizing the Universe with a Low Galaxy Ionizing Photon Escape Fraction, ApJ, 879, 36 (2019)
\bibitem{Graziani2020b} Graziani, L. et al., Cosmic archaeology with massive stellar black hole binaries , MNRASL, 495, Issue 1, pp.L81-L85 (2020)
\bibitem{Chon21} Chon, S. et al., Transition of the initial mass function in the metal-poor environments,  arXiv:2103.04997 (2021)
\bibitem{Chrimes20} Chrimes, A. et al., Binary population synthesis models for core-collapse gamma-ray burst progenitors,  MNRAS, 491, 3479 (2020)
\bibitem{Madau14} Madau, P. \& Dickinson, M., Cosmic Star-Formation History,  ARA\&A, 52, 415 (2014)
\bibitem {Atek15} Atek, H. et al., Are Ultra-faint Galaxies at $z = 6$--8 Responsible for Cosmic Reionization? Combined Constraints from the Hubble Frontier Fields Clusters and Parallels, ApJ,   814, 69  (2015)
\bibitem{Bouwens17} Bouwens, R.,  et al., Extremely Small Sizes for Faint {$z\sim$} 2-8 Galaxies in the Hubble Frontier Fields: A Key Input for Establishing Their Volume Density and UV Emissivity, ApJ, 843, 41(2017)
\bibitem{Vanzella21} Vanzella, E.,  et al., The MUSE Deep Lensed Field on the Hubble Frontier Field MACS J0416. Star-forming complexes at cosmological distances, A\&A, 646, A57 (2021)
\bibitem {Tanvir12} Tanvir, N. R. et al., Star Formation in the Early Universe: Beyond the Tip of the Iceberg, ApJ,   754,   46 (2012)
\bibitem {McGuire16} McGuire, J. et al., Detection of Three Gamma-ray Burst Host Galaxies at $z \sim 6$, ApJ, 825, 135 (2016)
\bibitem {Hartoog15} Hartoog, O. et al., VLT/X-Shooter spectroscopy of the afterglow of the Swift GRB 130606A. Chemical abundances and reionisation at $z \sim 6$, A\&A, 580, 139 (2015)
\bibitem {Fynbo09} Fynbo J.~P.~U.,  et al.,
Low-resolution spectroscopy of gamma-ray bursts optical afterglows: Biases in the SWIFT sample and characterization of the absorbers, ApJS, 185, 526 (2009)
\bibitem {Heintz18} Heintz, K. E.  et al., Highly ionized metals as probes of the circumburst gas in the natal regions of gamma-ray bursts, MNRAS,   479,   3456 (2018)
\bibitem{Urata} Urata, Y. \& Huang, K., GRB and Transient Sciences, http://ngvla.nao.ac.jp/researcher/memo/pdf/Urata.pdf
\bibitem{Runco21} Runco, J. N. et al., The MOSDEF survey: a comprehensive analysis of the rest-optical emission-line properties of $z \sim 2.3$ star-forming galaxies,  MNRAS, 502, 2600 (2021)
\bibitem {Maiolino08} Maiolino R., et al., AMAZE I. The evolution of the mass–metallicity relation at $z >3$, A\&A, 488, 463 (2008)
\bibitem {Moller13} M{\o}ller P. et al., Mass–metallicity relation from $z = 5$ to the present: evidence for a transition in the mode of galaxy growth at $z = 2.6$ due to the end of
sustained primordial gas infall, MNRAS, 430, 2680 (2013)
\bibitem{Laporte2018} Laporte, N. et al., Dust in the Reionization Era: ALMA Observations of a z = 8.38 Gravitationally Lensed Galaxy, ApJL, 837,  21 (2017)
\bibitem{Tamura2019} Tamura, Y. et al., Detection of the Far-infrared [O III] and Dust Emission in a Galaxy at Redshift 8.312: Early Metal Enrichment in the Heart of the Reionization Era, ApJ, 874, 27 (2019).
\bibitem{Faisst2020} Faisst, A. L. et al., ApJ Supplement Series,  247,  61  (2020)
\bibitem{McKinnon2017} McKinnon, R., et al., Simulating the dust content of galaxies: successes and failures , MNRAS, 468, 1505 (2017)
\bibitem{Aoyama2018} Aoyama, S., et al., Cosmological simulation with dust formation and destruction, MNRAS, 478, 4905 (2018)
\bibitem{Aoyama2017} Aoyama, S., et al., Galaxy simulation with dust formation and destruction, MNRAS, 466, 105 (2017)
\bibitem{Granato2020} Granato L., et al., Dust evolution in zoom-in cosmological simulations of galaxy formation, MNRAS, 503, 511 (2020)
\bibitem {Schady12} Schady,  P. et al., The dust extinction curves of gamma-ray burst host galaxies, A\&A,   537,   15 (2012)
\bibitem {Zafar18} Zafar, T. et al., X-shooting GRBs at high redshift: probing dust production history,  MNRAS,   480,  108 (2018)
\bibitem {Fynbo14} Fynbo, J.  P. U. et al., The mysterious optical afterglow spectrum of GRB 140506A at $z = 0.889$, A\&A,   572,   12 (2014)
\bibitem {Zafar11} Zafar, T. et al., The extinction curves of star-forming regions from $z = 0.1$ to 6.7 using GRB afterglow spectroscopy, ApJ,  735,   2 (2011)
\bibitem {Friis15} Friis, M. et al., The warm, the excited, and the molecular gas: GRB 121024A shining through its star-forming galaxy, MNRAS,   451,  167 (2015)
\bibitem{Noterdaeme17} Noterdaeme, P. et al., Discovery of a Perseus-like cloud in the early Universe. HI-to-H$_2$ transition, carbon monoxide and small dust grains at $z_{\rm abs} \approx 2.53$ towards the quasar J0000+0048, A\&A, 597, 82 (2017)
\bibitem{Balashev19} Balashev, S. A. et al., X-shooter observations of strong H$_2$-bearing DLAs at high redshift, MNRAS, 490, 2668 (2019)
\bibitem{Noterdaeme18} Noterdaeme, P. et al., Spotting high-$z$ molecular absorbers using neutral carbon. Results from a complete spectroscopic survey with the VLT, A\&A, 612, 58 (2018)
\bibitem {Heintz19} Heintz, K. E.  et al., New constraints on the physical conditions in H$_2$-bearing GRB-host damped Lyman-$\alpha$ absorbers, A\&A,   629,   131 (2019)
\bibitem {DeCia18} De Cia, A., et al., The cosmic evolution of dust-corrected metallicity in the neutral gas, A\&A,   611,   76 (2018)
\bibitem {Bolmer18} Bolmer, J. et al., Dust reddening and extinction curves toward gamma-ray bursts at $z > 4$, A\&A,   609,   62 (2018)
\bibitem{Campana07} Campana, S. et al., A Metal-rich Molecular Cloud Surrounds GRB 050904 at Redshift 6.3,  ApJ, 654, L17  (2007)
\bibitem {Tanvir19} Tanvir, N. R. et al., The fraction of ionizing radiation from massive stars that escapes to the intergalactic medium, MNRAS,   483,  5380 (2019)
\bibitem {Vielfaure20} Vielfaure, J.-B., et al., Lyman continuum leakage in faint star-forming galaxies at redshift $z = 3$--3.5 probed by gamma-ray bursts, A\&A,   641,   30 (2020)
\bibitem {Steidel18} Steidel, C. et al., The Keck Lyman Continuum Spectroscopic Survey (KLCS): The Emergent Ionizing Spectrum of Galaxies at $z \sim 3$, ApJ,   869, 123 (2018)
\bibitem {Liu16} Liu, C. et al., Dark-ages reionization and galaxy formation simulation - IV. UV luminosity functions of high-redshift galaxies, MNRAS,   462, 235  (2016)
\bibitem {Robertson13} Robertson, B. et al., New Constraints on Cosmic Reionization from the 2012 Hubble Ultra Deep Field Campaign, ApJ,   768,  71  (2013)
\bibitem{Totani06} Totani, T. et al., Implications for Cosmic Reionization from the Optical Afterglow Spectrum of the Gamma-Ray Burst 050904 at $z=6.3$, PASJ, 58, 485 (2006)
\bibitem{Chornock13} Chornock, R. et al., GRB 130606A as a Probe of the Intergalactic Medium and the Interstellar Medium in a Star-forming Galaxy in the First Gyr After the Big Bang, ApJ, 774, 26 (2013)
\bibitem {McQuinn08} McQuinn, M., et al., Probing the neutral fraction of the IGM with GRBs during the epoch of reionization, MNRAS,   388, 1101 (2008)
\bibitem{fuller20} Fuller, S. et al., Spectroscopically Confirmed Ly\,$\alpha$ Emitters from Redshift 5 to 7 behind 10 Galaxy Cluster Lenses,  ApJ, 896, 156 (2020)
\bibitem{mortlock11} Mortlock, D. et al., A luminous quasar at a redshift of $z = 7.085$,  Nature, 474, 616 (2011)

\bibitem {Bromm11} Bromm, V. \& Yoshida, N., The First Galaxies, AR A\&A,   49,  373  (2011)
\bibitem {Meszaros10} M\'esz\'aros, P. \& Rees, M.,  Population III Gamma-ray Bursts, ApJ,   715, 967  (2010)
\bibitem {Toma11} Toma, K., et al., \ Population III Gamma-ray Burst Afterglows: Constraints on Stellar Masses and External Medium Densities, \ ApJ, 731, 127  (2011)
\bibitem {Yoon15} Yoon, S-C. et al., Can Very Massive Population III Stars Produce a Super-Collapsar?, ApJ,   802, 16  (2015)
\bibitem {Burlon16} Burlon, D., et al., Gamma-ray bursts from massive Population-III stars: clues from the radio band, MNRAS,   459, 3356  (2016)
\bibitem{MB2014} Maio, U., Barkov, M.~V.\ Signatures of very massive stars: supercollapsars and their cosmological rate.\ MNRAS, 439, 3520–3525 (2014)
\bibitem {Ma15}  Ma, Q. et al., PopIII signatures in the spectra of PopII/I GRBs, MNRAS, 449, 3006 (2015)
\bibitem {Wang12}  Wang, F. Y., et al., Probing Pre-galactic Metal Enrichment with High-Redshift Gamma Ray Bursts, ApJ, 760, 27 (2012)
\bibitem{Ma17} Ma, Q. et al., Constraining the PopIII IMF with high-$z$ GRBs, MNRAS, 466, 1140 (2017)
\bibitem{Ciardi15} Ciardi, B. et al., Simulating the 21 cm forest detectable with LOFAR and SKA in the spectra of high-$z$ GRBs, MNRAS, 453, 101, (2015)
\bibitem{Maio2011} Maio, U., Khochfar, S., Johnson, J.~L., Ciardi, B. \ The interplay between chemical and mechanical feedback from the first generation of stars.\ MNRAS, 414, 1145–1157 (2011)
\bibitem {Amati02} Amati, L. et al., Intrinsic spectra and energetics of BeppoSAX Gamma-Ray Bursts with known redshifts,  A\&A,   390,  81 (2002)
\bibitem {Yonetoku04} Yonetoku, D. et al., Gamma-Ray Burst Formation Rate Inferred from the Spectral Peak Energy-Peak Luminosity Relation, ApJ,   609, 935 (2004)
\bibitem{Ghirlanda04b}The Collimation-corrected Gamma-Ray Burst Energies Correlate with the Peak Energy of Their $\nu F_{\nu}$ Spectrum, ApJ, 616, 331 (2004)
\bibitem{butler10} Butler, N. R. et al., The Cosmic Rate, Luminosity Function, and Intrinsic Correlations of Long Gamma-Ray Bursts,  ApJ, 711, 495 (2010)
\bibitem {Ghirlanda04} Ghirlanda, G. et al., Gamma-Ray Bursts: New Rulers to Measure the Universe, ApJ,   613, 13 (2004)
\bibitem {Amati08} Amati, L. et al., Measuring the cosmological parameters with the $E_{\rm p,i}$--$E_{\rm iso}$ correlation of gamma-ray bursts, MNRAS,   391, 577 (2008)
\bibitem{DiValentino20} Di Valentino, E., Melchiorri, A., Mena, O., Vagnozzi, S., Nonminimal dark sector physics and cosmological tensions PhRvD, 101, 063502 (2020) 
\bibitem{Lusso19} Lusso, E., et al., Tension with the flat $\Lambda$CDM model from a high-redshift Hubble diagram of supernovae, quasars, and gamma-ray bursts, A\&A, 628, L4 (2019)
bibitem{MV2015} Maio U., Viel M.,\ The first billion years of a warm dark matter universe, \ MNRAS, 446, 2760 (2015)
\bibitem{quintessence06} Maio, U. et al. \ Early structure formation in quintessence models and its implications for cosmic reionization from first stars.\ MNRAS, 373, 869–878 (2006) 
\bibitem{bulkflows11} Maio, U., Koopmans, L.~V.~E., Ciardi, B.\ The impact of primordial supersonic flows on early structure formation, reionization and the lowest-mass dwarf galaxies. \ MNRAS 412, L40–L44 (2011)
\bibitem{Maio2012} Maio, U., et al. \ Counts of high-redshift GRBs as probes of primordial non-Gaussianities.\ MNRAS 426, 2078–2088 (2012)






\end{thebibliography}


\end{document}